\documentclass{aastex63}
\usepackage{amsmath}

\shortauthors{A. Sicilia et al.}
\shorttitle{Stellar BHMF}

\begin{document}

\title{The Black Hole Mass Function Across Cosmic Times
\\ \small{I. Stellar Black Holes and Light Seed Distribution}}

\author[0000-0002-4515-3540]{Alex Sicilia}\affiliation{SISSA, Via Bonomea 265, 34136 Trieste, Italy}

\author[0000-0002-4882-1735]{Andrea Lapi}
\affiliation{SISSA, Via Bonomea 265, 34136 Trieste, Italy}\affiliation{IFPU - Institute for fundamental physics of the Universe, Via Beirut 2, 34014 Trieste, Italy}\affiliation{INFN-Sezione di Trieste, via Valerio 2, 34127 Trieste,  Italy}\affiliation{IRA-INAF, Via Gobetti 101, 40129 Bologna, Italy}

\author[0000-0003-3127-922X]{Lumen Boco}\affiliation{SISSA, Via Bonomea 265, 34136 Trieste, Italy}\affiliation{IFPU - Institute for fundamental physics of the Universe, Via Beirut 2, 34014 Trieste, Italy}

\author[0000-0003-0930-6930]{Mario Spera}\affiliation{SISSA, Via Bonomea 265, 34136 Trieste, Italy}\affiliation{IFPU - Institute for fundamental physics of the Universe, Via Beirut 2, 34014 Trieste, Italy}\affiliation{INFN-Sezione di Trieste, via Valerio 2, 34127 Trieste,  Italy}

\author[0000-0003-2654-5239]{Ugo N. Di Carlo}\affiliation{Physics and Astronomy Department Galileo Galilei, University of Padova, Vicolo dell'Osservatorio 3, 35122 Padova, Italy}\affiliation{
INFN-Sezione di Padova, Via Marzolo 8, 35131 Padova, Italy}

\author[0000-0001-8799-2548]{Michela Mapelli}\affiliation{Physics and Astronomy Department Galileo Galilei, University of Padova, Vicolo dell'Osservatorio 3, 35122 Padova, Italy}\affiliation{
INFN-Sezione di Padova, Via Marzolo 8, 35131 Padova, Italy}
\affiliation{INAF-OAPD, Vicolo dell'Osservatorio 5, 35122 Padova, Italy}

\author[0000-0001-8973-5051]{Francesco Shankar}\affiliation{Department of Physics and Astronomy, University of Southampton, Highfield SO17 1BJ, UK}

\author[0000-0002-5896-6313]{David M. Alexander}\affiliation{Department of Physics, Durham University, South Road, Durham, DH1 3LE, UK}

\author[0000-0002-7922-8440]{Alessandro Bressan}\affiliation{SISSA, Via Bonomea 265, 34136 Trieste, Italy}\affiliation{IFPU - Institute for fundamental physics of the Universe, Via Beirut 2, 34014 Trieste, Italy}

\author[0000-0003-1186-8430]{Luigi Danese}\affiliation{SISSA, Via Bonomea 265, 34136 Trieste, Italy}\affiliation{IFPU - Institute for fundamental physics of the Universe, Via Beirut 2, 34014 Trieste, Italy}

\begin{abstract}
This is the first paper in a series aimed at modeling the black hole (BH) mass function, from the stellar to the intermediate to the (super)massive regime. In the present work we focus on stellar BHs and provide an ab-initio computation of their mass function across cosmic times; we mainly consider the standard, and likely dominant production channel of stellar-mass BHs constituted by isolated single/binary star evolution. Specifically, we exploit the state-of-the-art stellar and binary evolutionary code \texttt{SEVN}, and couple its outputs with redshift-dependent galaxy statistics and empirical scaling relations involving galaxy metallicity, star-formation rate and stellar mass. The resulting relic mass function ${\rm d}N/{\rm d}V{\rm d}\log m_\bullet$ as a function of the BH mass $m_\bullet$ features a rather flat shape up to $m_\bullet\approx 50\, M_\odot$ and then a log-normal decline for larger masses, while its overall normalization at a given mass increases with decreasing redshift. We highlight the contribution to the local mass function from isolated stars evolving into BHs and from binary stellar systems ending up in single or binary BHs. We also include the distortion on the mass function induced by binary BH mergers, finding that it has a minor effect at the high-mass end. We estimate a local stellar BH relic mass density of $\rho_\bullet\approx 5\times 10^7\, M_\odot$ Mpc$^{-3}$, which exceeds by more than two orders of magnitude that in supermassive BHs; this translates into an energy density parameter $\Omega_\bullet\approx 4\times 10^{-4}$, implying that the total mass in stellar BHs amounts to $\lesssim 1\%$ of the local baryonic matter. We show how our mass function for merging BH binaries compares with the recent estimates from gravitational wave observations by LIGO/Virgo, and discuss the possible implications for dynamical formation of BH binaries in dense environments like star clusters. We address the impact of adopting  different binary stellar evolution codes (\texttt{SEVN} and \texttt{COSMIC}) on the mass function, and find the main differences to occur at the high mass end, in connection with the numerical treatment of stellar binary evolution effects. We highlight that our results can provide a firm theoretical basis for a physically-motivated light seed distribution at high redshift, to be implemented in semi-analytic and numerical models of BH formation and evolution. Finally, we stress that the present work can constitute a starting point to investigate the origin of heavy seeds and the growth of (super)massive BHs in high-redshift star-forming galaxies, that we will pursue in forthcoming papers.
\end{abstract}


\section{Introduction}\label{sec|intro}

The formation and evolution of black holes (BHs) in the Universe is one of the major issue to be addressed by the modern research in astrophysics and cosmology. In the mass range $m_\bullet\sim 5-150\, M_\odot$, BHs are originated from the final, often dramatic stages in the evolution of massive stars (possibly hosted in binary systems). These compact remnants can produce luminous X-ray binaries (e.g., Mapelli et al. 2010; Farr et al. 2011; Inoue et al. 2016), can constitute powerful sources of gravitational waves for ground-based detectors like the current LIGO/Virgo facility (e.g., Belczynski et al. 2010; Dominik et al. 2015; Spera et al. 2017, 2019; Boco et al. 2019; Abbott et al. 2021a,b), can possibly energize short gamma-ray bursts and associated kilonovas (e.g., Abbott et al. 2020, 2021c; Ackley et al. 2020; Gompertz et al. 2020), can inject strong energy inputs in the primeval Universe (e.g., Mirabel et al. 2011; Justham \& Schawinski 2012; Artale et al. 2015; Madau \& Fragos 2017; Lehmer et al. 2021), and can provide light seeds for the subsequent growth of more massive BHs (e.g., Madau et al. 2014; Volonteri et al. 2015; Lupi et al. 2016; Pacucci et al. 2017; Boco et al. 2020; Das et al. 2021). At the other end, in the range $M_\bullet\sim 10^6-10^{10}\, M_\odot$, supermassive BHs grow mainly by gaseous accretion, that energize the spectacular broadband emission of AGNs. Such an activity can have a profound impact on galaxy evolution (e.g., Alexander \& Hickox 2012; Lapi et al. 2014, 2018), as testified by the strict relationships between the relic BH masses and the physical properties of the hosts (e.g., Kormendy \& Ho 2013; Shankar et al. 2016, 2020; Zhu et al. 2021). The intermediate mass range $m_\bullet\sim 10^3-10^6\, M_\odot$ is the most uncertain. So far, only tentative evidence of these systems has been identified  (see Paynter et al. 2021). However, the chase is open in view of their astrophysical relevance. Most noticeably, they can provide heavy seeds for quick (super)massive BHs growth (e.g., Mayer \& Bonoli 2019; Boco et al. 2020), as it seems required by the puzzling observations of an increasing numbers of giant monsters $M_\bullet\gtrsim 10^9\, M_\odot$ when the age of the Universe was shorter than $\lesssim 0.8$ Gyr (e.g., Mortlock et al. 2011; Venemans et al. 2017; Banados et al. 2018). Moreover, such intermediate-mass BHs will constitute important targets for space-based gravitational wave detectors like LISA and DECIGO (see eLISA Consortium 2013; Kawamura et al. 2021; also Boco et al. 2021a; Barausse \& Lapi 2021).

One of the most fundamental quantity for demographic studies of the BH population is constituted by the relic mass function, namely the number density of BHs per comoving volume and unit BH mass, as a function of redshift. In the supermassive regime, where most of the BH mass is accumulated through gas accretion, this is usually determined from the AGN luminosity function via Soltan (1982)-type or continuity equation arguments (e.g., Small \& Blandford 1992; Salucci 1999; Yu \& Lu 2004; Merloni \& Heinz 2008; Kelly \& Merloni 2012; Aversa et al. 2015; Shankar et al. 2004, 2009, 2013, 2020), or from galaxy statistics and scaling relation among galactic and BH properties (e.g., Vika et al. 2009; Li et al. 2011; Mutlu-Pakdil et al. 2016). At the other end, the stellar BH mass function is largely unknown, and the most promising messenger to constrain it is constituted by the gravitational wave emission from coalescing binary BH systems (e.g., Kovetz et al. 2017; Perna et al. 2019; Tang et al. 2020; Ding et al. 2020); a first determination of the primary mass distribution for merging binary BHs has been established by the LIGO/Virgo collaboration (see Abbott et al. 2021a), although it depends somewhat on some assumptions (see Baxter et al. 2021). Therefore a theoretical grasp on the stellar BH mass function at different redshifts is of crucial importance. 

In the present work, we provide an ab-initio computation of the stellar BH relic mass function across cosmic times, by coupling the state-of-the-art stellar and binary evolutionary code \texttt{SEVN} to redshift-dependent galaxy statistics and empirical scaling relations involving metallicity, star-formation rate and stellar mass. Note that here we mainly consider the standard, and likely dominant production channel of stellar-mass BHs constituted by isolated single/binary star evolution, and defer to a future paper the treatment of other formation mechanisms like dynamical effects in dense star clusters (e.g., Miller \& Hamilton 2002; Alexander \& Natarajan 2014; Antonini et al. 2019), AGN disks (e.g., McKernan et al. 2012; Yang et al. 2019), hierarchical triples (e.g., Kimpson et al. 2016; Fragione et al. 2019) that are thought to produce corrections in the high-mass tail of the mass function, toward the intermediate-mass BH regime.

The plan of the paper is straightforward. In Sect. \ref{sec|theory} we introduce the theoretical background underlying our computation of the stellar BH relic mass function; specifically, we highlight the role of quantities related to galaxy formation (Sect. \ref{sec|galacticterm}) and to stellar evolution (Sect. \ref{sec|stellarterm}), and show how to include the effects of binary BH mergers (Sect. \ref{sec|mergrate}). In Sect. \ref{sec|results} we present our results concerning the stellar BH cosmic birthrate, the redshift-dependent stellar BH relic mass function and relic mass density; we also compare the primary mass distribution for merging BH binaries with the recent estimates from gravitational wave observations by LIGO/Virgo. In Sect. \ref{sec|discussion} we critically discuss how our results depend on the adopted binary stellar evolution code. Moreover, we estimate the effects of binary BH formation in dense environments like (young) star clusters. We also highlight the relevance of our computations in providing a light seed distributions for BH growth at high redshift. Finally, in Sect. \ref{sec|summary} we summarize our findings and outlook future developments.

Throughout this work, we adopt the standard flat $\Lambda$CDM cosmology (Planck Collaboration 2020) with rounded parameter values: matter density $\Omega_{\rm M}=0.3$, dark energy density $\Omega_{\Lambda}=0.7$, baryon density $\Omega_{\rm b}=0.05$, Hubble constant $H_0=100\, h$ km\ s$^{-1}$ Mpc$^{-1}$ with $h=0.7$. A Kroupa (2001) initial mass function (IMF) in the star mass range $m_\star\sim 0.1-150\, M_\odot$ and a value $Z_\odot\approx 0.015$ for the solar metallicity are adopted (see Caffau et al. 2011).

\section{Theoretical background}\label{sec|theory}

\begin{figure}
\centering
\includegraphics[width=\textwidth]{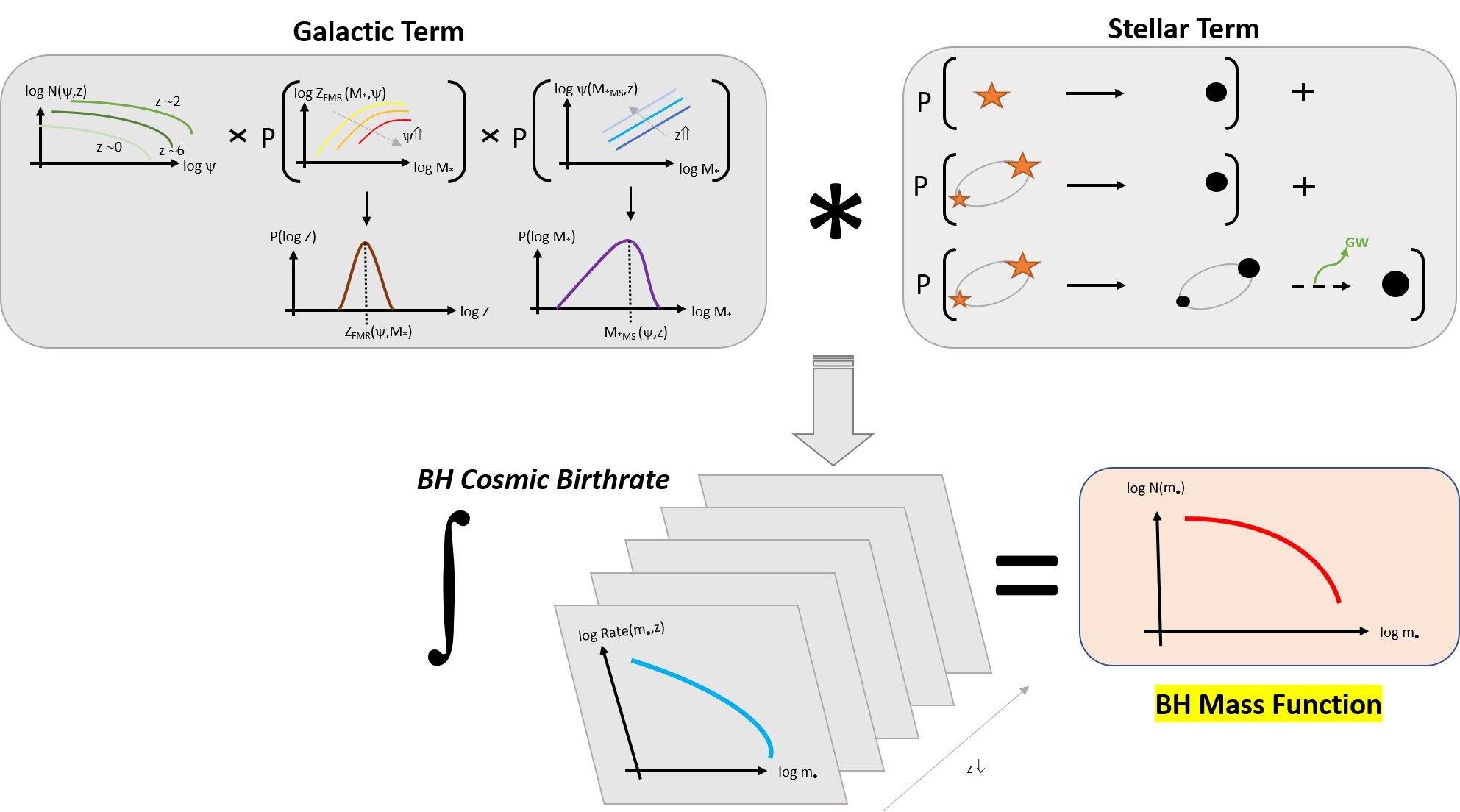}
\caption{Schematics showing the main steps to compute the stellar BH relic mass function (Eq. \ref{eq|massfunc}). This is obtained by integration over redshift of the BH cosmic birthrate (Eq. \ref{eq|birthrate}), which is in turn determined via the convolution of the galactic and the stellar terms. The galactic term (Eq. \ref{eq|galacticterm}) is computed by convolving the galaxy statistics (based on the SFR function) with the metallicity distribution of a galaxy at given SFR and stellar mass (assumed as a log-normal shape around the fundamental metallicity relation; see Eq. \ref{eq|probZ}) and with the stellar mass distribution of a galaxy at given SFR and redshift (built from the redshift-dependent main sequence of star-forming galaxies; see Eq. \ref{eq|probmstar}). The stellar term (Eq. \ref{eq|stellarterm}) is computed from the stellar and binary evolutionary code \texttt{SEVN} by summing up, for a given IMF, the probabilities per unit star-formed mass that: (i) a single star evolves into a BH remnant (single stellar evolution); (ii) a binary stellar system evolves into a single BH remnant or two BHs no longer bounded (failed binaries); (iii) a binary stellar system evolves into a binary BH (binaries), that may eventually coealesce into a single BH via emission of gravitational waves.}\label{fig|Schematic}
\end{figure}

We aim at deriving the stellar BH relic mass function, i.e., the number density of stellar BHs per unit comoving volume $V$ and remnant mass $m_\bullet$ accumulated down to redshift $z$
\begin{equation}\label{eq|massfunc}
\cfrac{{\rm d}N_{\bullet}}{{\rm d}V{\rm d}\log m_{\bullet}}(m_\bullet|>z)=\int_z^\infty{\rm d}z'\,\cfrac{{\rm d}t_{z'}}{{\rm d}z'}\,\cfrac{{\rm d}\dot N_{\bullet}}{{\rm d}V\,{\rm d}\log m_{\bullet}}(m_\bullet|z')\; ,
\end{equation}
and the related stellar BH relic mass density 
\begin{equation}\label{eq|BHMF_density}
\rho_\bullet(z) = \int{\rm d}\log m_\bullet\, m_\bullet\, \cfrac{{\rm d}N_{\bullet}}{{\rm d}V{\rm d}\log m_{\bullet}}(m_\bullet|z)\; .
\end{equation}
In the above Eq. (\ref{eq|massfunc}) the quantity $t_z$ is the cosmic time corresponding to redshift $z$ and ${\rm d}\dot N_\bullet/{\rm d}V\,{\rm d}\log m_{\bullet}$ is the stellar BH cosmic birthrate, i.e. the production rate of BH of given mass per unit comoving volume. In turn, the latter can be expressed as
\begin{equation}\label{eq|birthrate}
\cfrac{{\rm d}\dot N_\bullet}{{\rm d}V{\rm d}\log m_{\bullet}}(m_\bullet|z)=\int{\rm d}\log Z\,\cfrac{{\rm d}N_{\bullet}}{{\rm d}M_{\rm SFR}{\rm d}\log m_{\bullet}}(m_\bullet|Z)\,\cfrac{{\rm d}\dot{M}_{\rm SFR}}{{\rm d}V{\rm d}\log Z}(Z|z)
\end{equation}
where $Z$ is the metallicity and $M_{\rm SFR}$ is the star formed mass. The first and second factors in the integrand are usually referred to as the stellar and the galactic term, respectively (see Chruslinska \& Nelemens 2019; Boco et al. 2021b); we will now describe each of them in some detail, starting with the latter. The main steps toward the computation of the stellar BH relic mass function are sketched in Fig. \ref{fig|Schematic}.

\subsection{The galactic term}\label{sec|galacticterm}

The galactic term in Eq.~(\ref{eq|birthrate}) represents the cosmic SFR density per unit cosmic volume and metallicity; in other words, it constitutes the classic `Madau' plot (see Madau \& Dickinson 2014 and reference therein) sliced in metallicity bins, and as such is mainly related to galaxy formation and evolution. It may be estimated in various ways, starting from different galaxy statistics and empirical scaling laws. We compute it as
\begin{equation}\label{eq|galacticterm}
\cfrac{{\rm d}\dot{M}_{\rm SFR}}{{\rm d}V\,{\rm d}\log Z}(Z|z)=\int {\rm d}\log\psi\,\psi\,\cfrac{{\rm d}N}{{\rm d}V\,{\rm d}\log\psi}(\psi,z)\,\int {\rm d}\log M_\star\,\cfrac{{\rm d}p}{{\rm d}\log M_\star}(M_\star|\psi,z)\,\cfrac{{\rm d}p}{{\rm d}\log Z}(Z|M_\star,\psi)
\end{equation}
where $\psi$ is the SFR and $M_\star$ the stellar mass of a galaxy. Three ingredients enter into the above expression. The first is constituted by the redshift-dependent SFR functions ${\rm d}N/{\rm d}V\,{\rm d}\log\psi$, expressing the number of galaxies per unit cosmological volume and SFR bin. For these we adopt the determination by Boco et al. (2021, their Fig. 1;  for an analytic Schechter fit see Eq. 2 and Table 1 in Mancuso et al. 2016a) derived from an educated combination of the dust-corrected UV (e.g., Oesch et al. 2018; Bouwens et al. 2020), IR (e.g., Gruppioni et al. 2020; Zavala et al. 2021), and radio (e.g. Novak 2017; Ocran 2020) luminosity functions, appropriately converted in SFR (see Kennicutt \& Evans 2012). 

The second ingredient is the probability distribution of stellar mass at given SFR and redshift:
\begin{equation}\label{eq|probmstar}
\cfrac{{\rm d}p}{{\rm d}\log M_\star}(M_\star|\psi,z)\propto
\begin{cases}
M_\star & M_\star<M_{\star,\rm MS}(\psi,z)\\
M_{\star,\rm MS}\,\exp{\left\{-\cfrac{[\log M_\star-\log M_{\star,\rm MS}(\psi,z)]^2}{2\,\sigma_{\log M_\star}^2}\right\}} & M_\star\geq M_{\star,\rm MS}(\psi,z)
\end{cases}
\end{equation}
where $M_{\star,\rm MS}(\psi,z)$ is the observed redshift-dependent galaxy main sequence with log-normal scatter $\sigma_{\log M_\star}\approx 0.2$ dex (we adopt the determination by Speagle et al 2014; for an anlytic fit, see their Eq. 28). The main sequence is a relationship between SFR and stellar mass followed by the majority of star-forming galaxies, apart from some outliers located above the average SFR at given stellar mass (see Daddi et al. 2007; Sargent et al. 2012; Rodighiero et al. 2011, 2015; Speagle et al. 2014; Whitaker et al. 2014; Schreiber et al. 2015; Caputi et al. 2017; Bisigello et al. 2018; Boogaard et al. 2018). The expression in Eq. (\ref{eq|probmstar}) holds for an approximately constant SFR history, which is indicated both by  in-situ galaxy formation scenarios (see Mancuso et al. 2016b; Pantoni et al. 2019; Lapi et al. 2020) and by observations of ETG progenitors (that have on overage slowly rising star formation history with typical duration of $\lesssim 1$ Gyr; see Papovich et al. 2011; Smit et al. 2012; Moustakas et al. 2013; Steinhardt et al. 2014; Cassar\'a et al. 2016; Citro et al. 2016) and late-type galaxies (that have on the average slowly declining star formation history over long timescale of several Gyrs; e.g., see Chiappini et al. 1997; Courteau et al. 2014; Pezzulli \& Fraternali 2016; Grisoni et al. 2017). In this vein, off-main sequence objects can be simply viewed as galaxies caught in an early evolutionary stage, that are still accumulating their stellar mass (which grows almost linearly with time for a constant SFR), and are thus found to be preferentially located above the main sequence or, better, to the left of it. As time goes by and the stellar mass increases, the galaxy moves toward the average main sequence relationship, around which it will spend most of its lifetime before being quenched due to gas exhaustion or feedback processes. 

The third ingredient is the probability distribution of metallicity at given stellar mass and SFR
\begin{equation}\label{eq|probZ}
\cfrac{{\rm d}p}{{\rm d}\log Z}(Z|M_\star,\psi)\propto \exp{\left\{-\cfrac{[\log Z-\log Z_{\rm FMR}(M_\star,\psi)]^2}{2\,\sigma^2_{\log Z}}\right\}}
\end{equation}
that is assumed to be log-normal with mean $\log Z_{\rm FMR}(M_\star,\psi)$ and scatter $\sigma_{\log Z}\approx 0.15$ dex provided by the observed fundamental metallicity relation (we adopt the determination by Mannucci et al. 2011; for an analytic fit, see their Eq. 2). This is a relationship among metallicity, stellar mass and SFR of a galaxy which is found to be closely independent of redshift at least out to $z\lesssim 4$ (see Mannucci et al. 2010, 2011; Hunt et al. 2016; Curti et al. 2020; Sanders et al. 2021). In the aforementioned in-situ galaxy formation scenarios (see Pantoni et al. 2019; Lapi et al. 2020; Boco et al. 2021b), the fundamental metallicity relation naturally stems from: an early, rapid increase of the metallicity with galactic age, which for a roughly constant star formation history just amounts to the ratio $M_\star/\psi$; a late-time saturation of the metallicity to a mass-dependent value which is determined by the balance among metal dilution, astration, removal by feedback, and partial restitution by stellar evolution processes and galactic fountains (if present). 

\begin{figure}
\centering
\includegraphics[width=\textwidth]{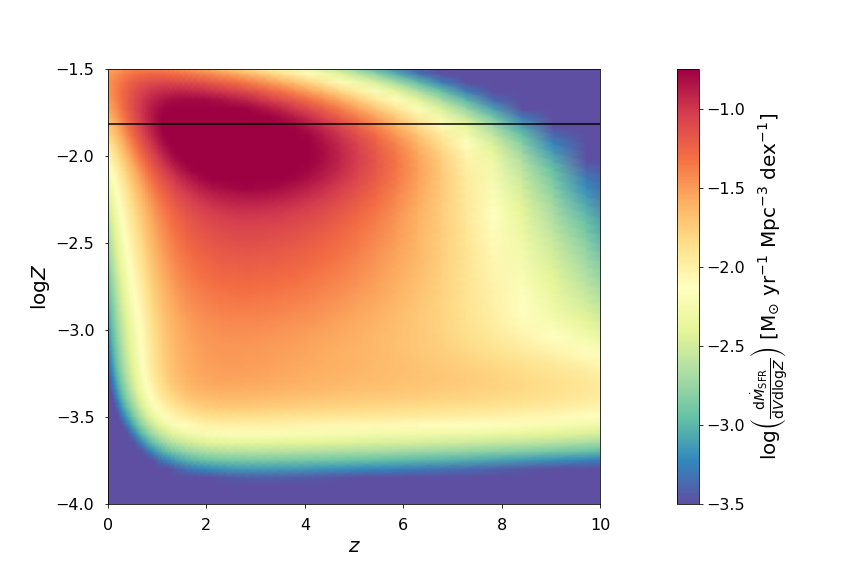}
\caption{The galactic term ${\rm d}\dot M_\star/{\rm d}V\,{\rm d}\log Z$ of Eq. (\ref{eq|galacticterm}) expressing the amount of SFR (color-coded) per unit comoving volume as a function of redshift $z$ (on $x$-axis) and of metallicity $Z$ (on $y$-axis). The black horizontal line highlights the solar value $Z_\odot$.}\label{fig|Galterm}
\end{figure}

In Fig. \ref{fig|Galterm} we illustrate the galactic term given by Eq. (\ref{eq|galacticterm}). Specifically, we have  color-coded the cosmic SFR density ${\rm d}\dot M_{\rm SFR}/{\rm d}V\,{\rm d}\log Z$ per unit comoving volume as a function of redshift $z$ and metallicity $\log Z$. Most of the cosmic star formation occurs around redshift $z\sim 2-3$ at rather high metallicity, close to solar values. The metallicity at which most star formation takes place decreases with redshift, but very mildly out to $z\sim 4-5$. However, we note that at any redshift there is a long, pronounced  tail of star formation occurring at low metallicities, down to $0.01\, Z_\odot$. We discuss possible variations of the prescriptions entering the galactic term in Sect. \ref{sec|discussion_modeling}. 

\subsection{The stellar term}\label{sec|stellarterm}

The stellar term in Eq.~(\ref{eq|birthrate}) represents the number of BHs formed per unit of star formed mass and remnant mass; as such, it is related to the evolution of isolated or binary stars, and must be evaluated via detailed simulations of stellar astrophysics. The stellar term can be split in various contributions
\begin{equation}\label{eq|stellarterm}
\cfrac{{\rm d}N_\bullet}{{\rm d}M_{\rm SFR}{\rm d}\log m_{\bullet}}(m_\bullet|Z)=\cfrac{{\rm d}N_{\star\rightarrow \bullet}}{{\rm d}M_{\rm SFR}{\rm d}\log m_{\bullet}}(m_\bullet|Z)+\cfrac{{\rm d}N_{\star\star\rightarrow \bullet}}{{\rm d}M_{\rm SFR}{\rm d}\log m_{\bullet}}(m_\bullet|Z)+\sum_{i=1,2}\cfrac{{\rm d}N_{\star\star\rightarrow \bullet\bullet}}{{\rm d}M_{\rm SFR}{\rm d}\log m_{\bullet,i}}(m_\bullet|Z)
\end{equation}
the first comes from the evolution of isolated, massive stars that evolve into BHs at the end of their life (hereafter referred to as `single stellar evolution'); the second comes from stars that are originally in binary systems but end up as an isolated BH because one of the companion has been ejected or destroyed or cannibalized (hereafter `failed BH binaries'); the third comes from stars in binary systems that evolve into a binary BH with primary mass $m_{\bullet,1}$ and secondary mass $m_{\bullet,2}$ (hereafter `binaries'). All these terms are strongly dependent on metallicity $Z$, since this quantity affects the efficiency of the various processes involved in stellar and binary evolution, like mass loss rates, mass transfers, core-collapse physics, etc. Note that other crucial ingredients implicitly entering the above terms are the IMF $\phi(m_\star)$, i.e. the distribution of star masses $m_\star$ per unit mass formed in stars, and the binary fraction $f_{\star\star}$, i.e. the mass fraction of stars originally born in binary systems; both are rather uncertain quantities. As a reference we adopt a universal (i.e., equal for every galaxies at any cosmic time) Kroupa (2001) IMF and a binary mass fraction $f_{\star\star}\approx 0.5$ constant with the star mass $m_\star$. We discuss the impact of these choices on our results in Sect. \ref{sec|discussion_modeling}.

To compute the stellar term, we exploit the outcomes of the \texttt{SEVN} stellar and binary evolution code, that provides directly each of the above contributions (see Spera et al. 2019 for details). However, to add some more grasp on the physics involved, we can provide a handy approximation for the first term referring to isolated stars; this writes as
\begin{equation}\label{eq|singlestar}
\cfrac{{\rm d}N_{\star\rightarrow \bullet}}{{\rm d}M_{\rm SFR}{\rm d}\log m_{\bullet}}(m_\bullet|Z) = (1-f_{\star\star}) \int{\rm d}m_\star\, \phi(m_\star)\,\cfrac{{\rm d}p_{\star\rightarrow \bullet}}{{\rm d}\log m_\bullet}(m_\bullet|m_\star,Z)
\end{equation}
where $f_{\star\star}$ is the binary fraction, $\phi(m_\star)$ is the IMF, and
\begin{equation}\label{eq|probmrem}
\cfrac{{\rm d}p_{\star\rightarrow \bullet}}{{\rm d}\log m_\bullet}(m_\bullet|m_\star,Z)\propto \exp{\left\{-\cfrac{[\log m_\bullet-\log m_{\star\rightarrow \bullet}(m_\star,Z)]^2}{2\,\sigma^2_{\log m_\bullet}}\right\}}
\end{equation}
is an approximately log-normal distribution centered around the metallicity-dependent relationship $m_{\star\rightarrow \bullet}(m_\star,Z)$ between the remnant mass and the initial zero-age main sequence star mass, with dispersion $\sigma_{\log m_\bullet}\approx 0.15$ dex (see Spera et al. 2017; Boco et al. 2019). Unfortunately, for the other contributions of the stellar term related to binary evolution, an analogous analytic approximation is not viable, so one must trust the outputs of the \texttt{SEVN} code. We discuss the impact of adopting different prescriptions and codes for the computation of the stellar term in Sect. \ref{sec|discussion_modeling}. 

In Fig. \ref{fig|Stellarterm} we illustrate the stellar term of Eq. (\ref{eq|stellarterm}), split in its various contributions. Specifically, we have color-coded the distributions ${\rm d}N_\bullet/{\rm d}M_{\rm SFR}\,{\rm d}\log m_\bullet$ per unit star mass formed $M_{\rm SFR}$ and BH remnant mass $m_\bullet$. The top left panel refers to isolated stars evolving in BHs. It is seen that a roughly constant number of remnants with masses $m_\bullet\sim 5-30\, M_\odot$ is produced per logarithmic BH mass bin at any metallicity. Then there is a peak around $m_\bullet\sim 30-60\, M_\odot$, with the larger values applying to metal-poorer conditions, where stellar winds are not powerful enough to substantially erode the stellar envelope before the final collapse. Finally, the distribution rapidly falls off for larger $m_\bullet\gtrsim 50-60\, M_\odot$ even at low metallicity\footnote{Note that adopting an IMF with an upper star mass limit $m_\star\sim 150\, M_\odot$ avoids dealing with the formation of high-mass remnants $m_\bullet\gtrsim 100\, M_\odot$ by direct collapse.}, due to the presence of pair-instability and pulsational pair-instability supernovae (see Woosley et al. 2002; Belczynski et al. 2016; Woosley 2017; Spera et al. 2017). 

The top right panel refers to binary stellar systems failing to form a compact binary, and evolving instead into isolated BHs; this may happen because one of the progenitor star has been ejected far away or destroyed or cannibalized during binary stellar evolution. 
The distribution of failed BH binaries differs substantially from single stellar evolution, being skewed toward more massive BHs, and  with an appreciable number of remnants of mass $m_\bullet\sim 50-160\, M_\odot$ produced especially at low metallicities. Such massive BH remnants are mainly formed when two (non-degenerate) companion stars merge during a common envelope phase, possibly leaving then a big BH remnant. Finally, note that at high metallicity a non-negligible fraction of BHs in this channel has formed after the low-mass companion star had been ejected far away or destroyed by stellar winds and/or supernova explosions (see Spera et al. 2019 for details). The bottom left panel refers to binary stellar systems evolving into binary BHs; note that the distribution of primary and secondary BHs in the final configuration have been summed over. Although the overall number of binary BHs is substantially lower than the single BHs originated from the other two channels, most of them has a very similar mass spectrum; these remnants have formed from binary stars that underwent minor mass transfer episodes. However, at low metallicity stellar winds are reduced and hence the mass exchanged or lost during binary evolution may be significantly larger, implying a more extended tail toward masses $m_\bullet\lesssim 100\, M_\odot$ with respect to single stellar evolution. Finally, the bottom right panel illustrates the sum of all the previous formation channels.

\subsection{Binary BH mergers}\label{sec|mergrate}

Tight BH binaries may be able to progressively lose their energy via gravitational wave emission and to merge in a single, more massive BH. The cosmic merging rate of binary BHs can be computed as 
\begin{equation}\label{eq|mergrate}
\cfrac{{\rm d}\dot{N}_{\bullet\bullet\rightarrow \bullet}}{{\rm d}V\,{\rm d}\log m_{\bullet,i}}(m_\bullet|z)=\int{\rm d} t_{\rm d}\int{\rm d}\log Z\,\cfrac{{\rm d}N_{\star\star\rightarrow \bullet\bullet\rightarrow \bullet}}{{\rm d}M_{\rm SFR}{\rm d}\log m_{\bullet,i}}(m_\bullet|Z)\,\cfrac{{\rm d}p}{{\rm d}t_{\rm d}}(t_{\rm d}|Z)\,\cfrac{{\rm d}\dot{M}_{\rm SFR}}{{\rm d}V{\rm d}\log Z}(Z|z_{t-t_{\rm d}})~~~~~~~~~i=1,2
\end{equation}
where $z_t$ is the redshift corresponding to a cosmic time $t$, and $t_{\rm d}$ is the time delay between the formation of the progenitor binary and the merger of the compact binary. The integrand in the expression above is the product of three terms. The rightmost one is the galactic term that must be computed at a time $t-t_{\rm d}$. The middle one represents the probability distribution of delay times, possibly dependent on metallicity. The leftmost one is the stellar term that represents the number of binary stellar systems first evolving in a compact binary and then being able to merge within the Hubble time, per unit of formed star mass and compact remnant mass (primary $m_{\bullet,1}$ and secondary $m_{\bullet,2}$); this is a fraction of the third term on the right hand side of Eq.~(\ref{eq|stellarterm}). 
The merger rate is then given by convolution of such a specific stellar term with the galactic term, weighted by the time delay distribution.

As a consequence of binary BH mergers, the stellar BH relic mass function may be somewhat distorted, since a number of low mass BHs are shifted to larger masses. This amounts to apply a merging correction to the original cosmic birthrate ${\rm d}\dot N/{\rm d}V\,{\rm d}m_\bullet$, that can be computed as follows\footnote{For the sake of simplicity in Eq. (\ref{eq|birthratecorr}) we are neglecting the small amount of mass lost as gravitational waves during the coalescence, since we checked that for the purpose of this computation it is practically irrelevant.}
\begin{equation}\label{eq|birthratecorr}
\begin{aligned}
\cfrac{{\rm d}\dot N_{\bullet,\rm mergcorr}}{{\rm d}V{\rm d} m_{\bullet}}(m_\bullet|z) &= \cfrac{{\rm d}\dot N_{\bullet}}{{\rm d}V{\rm d} m_{\bullet}}(m_\bullet|z)  + \cfrac{1}{2}\,\int_{m_\bullet/2}^{m_\bullet}{\rm d}m_{\bullet,1}\, \cfrac{{\rm d}\dot N_{\bullet\bullet\rightarrow \bullet}}{{\rm d}V\,{\rm d} m_{\bullet,1}}(m_{\bullet,1}|z)\,\cfrac{\cfrac{{\rm d}\dot{N}_{\bullet\bullet\rightarrow \bullet}}{{\rm d}V\,{\rm d} m_{\bullet,2}}(m_\bullet-m_{\bullet,1}|z)}{\int_0^{m_{\bullet,1}}{\rm d}m_{\bullet,2}\, \cfrac{{\rm d}\dot{N}_{\bullet\bullet\rightarrow \bullet}}{{\rm d}V\,{\rm d} m_{\bullet,2}}(m_{\bullet,2}|z)}+\\
\\
& + \cfrac{1}{2}\int_{0}^{m_\bullet/2}{\rm d}m_{\bullet,2}\, \cfrac{{\rm d}\dot N_{\bullet\bullet\rightarrow \bullet}}{{\rm d}V\,{\rm d} m_{\bullet,2}}(m_{\bullet,2}|z)\,\cfrac{\cfrac{{\rm d}\dot{N}_{\bullet\bullet\rightarrow \bullet}}{{\rm d}V\,{\rm d} m_{\bullet,1}}(m_\bullet-m_{\bullet,2}|z)}{\int_{m_{\bullet,2}}^\infty{{\rm d} m_{\bullet,1}}\,\cfrac{{\rm d}\dot{N}_{\bullet\bullet\rightarrow \bullet}}{{\rm d}V\,{\rm d} m_{\bullet,1}}(m_{\bullet,1}|z)}- \sum_{i=1,2}\, \cfrac{{\rm d}\dot{N}_{\bullet\bullet\rightarrow \bullet}}{{\rm d}V\,{\rm d} m_{\bullet,i}}(m_\bullet|z)\; ;
\end{aligned}
\end{equation}
in the positive terms on the right hand side, the normalizations of the rates have been chosen so as to ensure mass conservation and self-consistency of the integrated merger rates. 

We caveat that additional distorsions of the relic BH mass function may be induced by dynamical effects and/or hierarchical mergers in dense environments (see Sect. \ref{sec|discussion_dynamics} for a preliminary discussion), by mass growth due to accretion in X-ray binaries or in AGN disks, etc.; these processes are expected to be relevant especially at the high mass end, toward the intermediate-mass BH regime.

\section{Results}\label{sec|results}

In Fig. \ref{fig|BHMF_rate} we illustrate the stellar BH cosmic birthrate of Eq. (\ref{eq|birthrate}) as a function of redshift $z$ and remnant mass $m_\bullet$. Specifically, we have color-coded the birthrate ${\rm d}\dot N_\bullet/{\rm d}V\,{\rm d}\log m_\bullet$ per unit comoving volume $V$ and BH mass $m_\bullet$. At $z\sim 1-4$ most of the BHs are formed with a rather flat distribution in $\log m_\bullet$ for $m_\bullet\sim 5-50\, M_\odot$, though there is a non-negligible tail up to $m_\bullet\lesssim 160\, M_\odot$ due to stellar mergers (see previous section for details). Moving toward high redshift $z\gtrsim 4$ the mass distribution becomes skewed toward masses $m_\bullet\sim 30-50\, M_\odot$ since these are preferentially produced in the low metallicity environments, while the tail for $m_\bullet\gtrsim 50\, M_\odot$ tends to be reduced because of the decrease in the number density of star-forming galaxies with appreciable SFR.

\begin{figure}
\centering
\includegraphics[width=\textwidth]{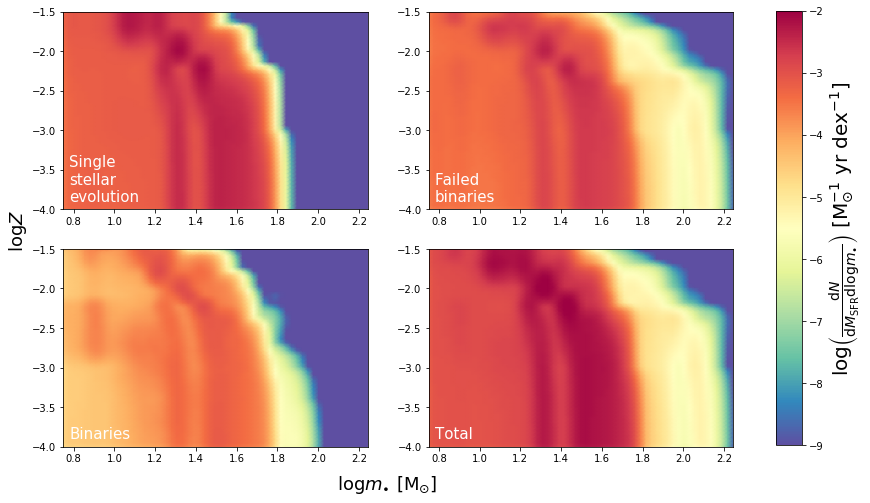}
\caption{The stellar term ${\rm d}N_\bullet/{\rm d}M_{\rm SFR}\,{\rm d}\log m_\bullet$ of Eq. (\ref{eq|stellarterm}) expressing the number of BHs per unit star-formed mass (color-coded) as a function of BH mass $m_\bullet$ (on $x$-axis) and of metallicity $Z$ (on $y$-axis). Different panels refer to: isolated stars evolving into single BH (top left); binary stars failing to form a compact binary and instead originating a single BH (top right); binary stars evolving in binary BHs (bottom left); summation of these contributions (bottom right).}\label{fig|Stellarterm}
\end{figure}

\begin{figure}
\centering
\includegraphics[width=\textwidth]{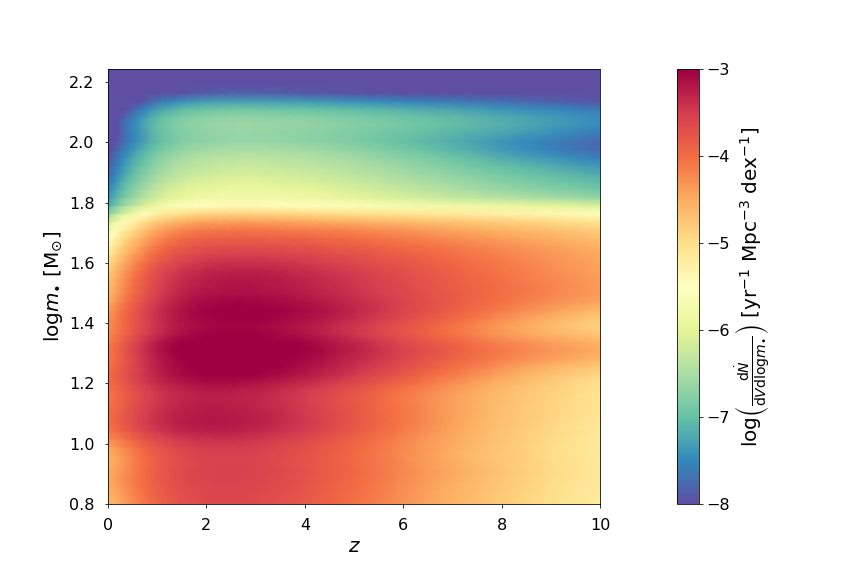}
\caption{The stellar BH cosmic birthrate ${\rm d}\dot N/{\rm d}V\,{\rm d}\log m_\bullet$ (color-coded) as a function of redshift $z$ (on $x$-axis) and BH mass $m_\bullet$ (on $y$-axis).}\label{fig|BHMF_rate}
\end{figure}

In Fig. \ref{fig|BHMF} we illustrate the stellar BH relic mass function ${\rm d}N_\bullet/{\rm d}V\,{\rm d}\log m_\bullet$ of Eq. (\ref{eq|massfunc}) as a function of the remnant mass $m_\bullet$ for different redshifts $z\sim 0-10$. At given redshift, the mass function features a roughly constant behavior for $m_\bullet\sim 5-50\, M_\odot$, followed by a quite steep decline for $m_\bullet\gtrsim 50\, M_\odot$. Noticeably, there are bumps at around $m_\bullet\sim 20\, M_\odot$, $m_\bullet\sim 30-50\, M_\odot$ and $m_\bullet\sim 120\, M_\odot$, that are more pronounced at high redshift (where metallicity is smaller) and progressively washed out toward the local Universe. Such features reflect regions of higher BH numbers that are evident in the stellar term of Fig. \ref{fig|Stellarterm} (in the form of almost vertical darker strips). Specifically, the first two originate from the detailed shape of the metallicity-dependent BH mass spectrum from single stellar evolution, while the third one mainly depends on binary evolution effects.

The mass function increases for decreasing redshift, quite rapidly down to $z\sim 2-3$ and then more mildly toward $z\sim 0$. Smoothing out minor features, the mass function can be analytically rendered in the range $m_\bullet\sim 5-160\, M_\odot$ via a Schechter$+$Gaussian shape (Saunders et al. 1990)
\begin{equation}\label{eq|fits}
    \frac{{\rm d}N}{{\rm d}V\,{\rm d}\log m_\bullet}\simeq \mathcal{N}\, \left(\frac{m_\bullet}{\mathcal{M}_\bullet}\right)^{1-\alpha}\, e^{-m_\bullet/\mathcal{M}_\bullet}+\mathcal{N}_G\,\cfrac{1}{\sqrt{2\pi\sigma_G^2}}\,e^{-(\log m_\bullet-\log  \mathcal{M}_{\bullet,G})^2/2\,\sigma_G^2}\; ;  
\end{equation}
the Schechter function has normalization $\mathcal{N}$, low-mass power-law index $\alpha$ and characteristic mass $\mathcal{M}_\bullet$, while the Gaussian (log-normal) has normalization $\mathcal{N}_G$, mean $\mathcal{M}_{\bullet,G}$ and dispersion $\sigma_G$. 
The optimal values of these parameters have been determined via a Levenberg-Marquardt least-squares fit to the numerical mass function for $m_\bullet\sim 5-160\, M_\odot$; the results for representative redshifts in the range $z\sim 0-10$ are reported in Table 1.

In Fig. \ref{fig|BHMF_split} we highlight the contribution of the different stellar evolution channels to the stellar BH relic mass function at $z\sim 0$ and at $z\sim 10$. In the range $m_\bullet\sim 5-50\, M_\odot$ the single stellar evolution and the failed BH binaries channels are very similar and dominates over BH binaries. 
For $m_\bullet\gtrsim 50\, M_\odot$ the single stellar evolution contribution sharply dies (due to the mass gap from pair-instability SNe, see Sect. \ref{sec|stellarterm}) and the binary BH channel abruptly decreases (due to mass loss in common envelope phase, see Sect. \ref{sec|stellarterm}), while the contribution from failed BH binaries dominates largely. Such a behavior in the relative contributions is basically independent of redshift.

In Fig. \ref{fig|BHMF_density} we illustrate the stellar BH relic mass density $\rho_\bullet(z)$ of Eq. (\ref{eq|BHMF_density}) as a function of redshift $z$. The mass density increases quite steeply toward smaller redshifts, changing from values $\lesssim 10^5\, M_\odot$ Mpc$^{-3}$ at $z\sim 10$ to  $10^7\, M_\odot$ Mpc$^{-3}$ around $z\sim 2-3$ and then saturating for lower redshifts up to $5\times 10^7\, M_\odot$ Mpc$^{-3}$ at $z\sim 0$. The evolution with redshift plainly follows that of the mass density $\rho_\star=(1-R)\,\int_0^z{\rm d}z'\int{\rm d}\log \psi\, \psi\, ({\rm d}N/{\rm d}V\,{\rm d}\log \psi)$ in stars within galaxies, that has been computed by integrating the SFR functions and taking into account the recycling fraction $R\approx 0.45$; this corresponds to a local energy density $\Omega_\star\approx 5\times 10^{-3}$, consistent with classic estimates (e.g., Fukugita \& Peebles 2004). On the gross average, we find a ratio of the mass densities $\rho_\bullet/\rho_\star\approx 10\%$; we have checked that this mainly reflects the fraction of stars in the Kroupa IMF with mass $m_\star\gtrsim 20\, M_\odot$ originating BH remnants, appropriately lowered by mass loss in stellar winds and by binary evolution effects. The local density in stellar mass BHs is substantially larger than that in supermassive BHs with masses $M_\bullet\sim 10^6-10^{10}\, M_\odot$, which amounts to $\sim 1-2\times 10^5\, M_\odot$ Mpc$^{-3}$ (see Shankar et al 2020; see also Kormendy \& Ho 2013; Aversa et al. 2015), implying that most of the BH mass in the local Universe is of stellar origin. The ratio of the local mass densities in supermassive to stellar mass BHs is around $\sim 2\times 10^{-3}$. Finally, we estimate a local BH energy density $\Omega_\bullet\approx 4\times 10^{-4}$, corresponding to a BH-to-baryon ratio $\Omega_\bullet/\Omega_b\lesssim 10^{-2}$, i.e. the total mass in stellar BHs amounts to $\lesssim 1\%$ of the local baryonic matter.

For the sake of completeness, in Fig. \ref{fig|BHMF_density} we illustrate the contributions to the BH mass density from different mass ranges: $m_\bullet\lesssim 20\, M_\odot$, $m_\bullet\sim 20-50\, M_\odot$ and $m_\bullet\gtrsim 50\, M_\odot$. As expected from the shape of the mass function, most of the mass density is contributed by BHs with mass in the interval $m_\bullet\sim 20-50\, M_\odot$. We also note that the mass density of BHs with masses $m_\bullet \lesssim 20\, M_\odot$ tends to decline more steeply than the total one toward high-redshift, because the production of such small mass BHs is disfavored in the lower-metallicity environment of high-$z$ galaxies; the opposite holds for the mass density of BHs with larger masses $m_\bullet\gtrsim 50\, M_\odot$.

In Fig. \ref{fig|BHMF_merging} we evaluate the impact of binary BH mergers on the relic mass function. In the top left panel we have color-coded the stellar term $\sum_{i=1,2}\,{\rm d}N_{\star\star\rightarrow\bullet\bullet\rightarrow \bullet}/{\rm d}M_{\rm SFR}\, {\rm d}\log m_{\bullet,i}$ representing the number of binary stellar systems that first evolve in a BH binary and then are able to merge within the Hubble time, as function of metallicity $Z$ and BH mass $m_\bullet$. The remnants contributing to this term are actually a sub-sample of those in the binary stellar term of Fig. \ref{fig|Stellarterm} (bottom left panel). It is evident that the number of merging binary BHs is a small fraction of the total, while its distribution lacks the high-mass tail. This is because to form a tight compact binary, the progenitor stars must have undergone a substantial phase of common envelope, and an ensuing mass loss/envelope ejection (if not, the two progenitors merge prematurely and a single BH remnant is left). The end-product is a binary which can be sufficiently tight to merge within an Hubble time, but typically made of BHs that cannot be more massive than $m_\bullet\lesssim 40-50\, M_\odot$ (see Giacobbo \& Mapelli 2018; Spera et al. 2019).

In the top right panel we have color-coded the cosmic merging rate $\sum_{i=1,2}\,{\rm d}\dot N_{\bullet\bullet\rightarrow \bullet}/{\rm d}V\, {\rm d}\log m_{\bullet,i}$ constituting the destruction term in Eq. (\ref{eq|mergrate}), as a function of BH mass $m_\bullet$ and redshift $z$. It is seen that most binary BH mergers occur for redshift $z\sim 1-5$ in the mass ranges $m_\bullet\sim 15-40\, M_\odot$ and $m_\bullet\sim 5-8\, M_\odot$. In the bottom panel, we show the stellar BH relic mass function at $z\sim 0$ with and without the correction for binary BH mergers; the solid lines is the total mass function, while the dashed lines are the contribution from BH binaries. As expected the main changes are at the high-mass end, where an appreciable number of BHs in binaries is shifted toward larger masses by the merging process; however, the net effect on the total mass function is minor. 

The contribution to the stellar BH mass function from merging binary BHs can be probed via gravitational wave observations. Recently, the LIGO/Virgo collaborations (Abbott et al. 2021a; see also Baxter et al. 2021) has estimated the primary mass distribution for BH binaries that coalesce around $z\approx 0$; given the quite small cosmological volume probed by the current gravitational wave detectors, this approximately corresponds to the merger rate integrated in a narrow redshift interval $\Delta z\lesssim 0.5$. The expectation from this work is illustrated as a red solid line in Fig. \ref{fig|BHMF_primary}, and compared with the estimates by Abbott et al. (2021a; blue shaded area) for a powerlaw$+$peak model and by Baxter et al. (2021; orange shaded area) for their astrophysically-motivated model (for both determinations the $68\%$ and $95\%$ credible intervals are represented with dark and light shades). Our result is in remarkable agreement with these estimates up to $40\, M_\odot$. However, the observed primary mass distribution declines gently for $m_\bullet\gtrsim 40\, M_\odot$ out to $m_\bullet\sim 80-100\, M_\odot$ while our model dies off, since stellar evolution effects hinder the presence of very massive BHs in coalescing binaries. This occurs mainly for two reasons (see also Sect. \ref{sec|results}): the mass gap $m_\bullet\sim 50-120\, M_\odot$ for the production of BH due to pair-instability and pulsational pair-instability supernovae; (ii) the substantial mass loss during the common-envelope phase needed to produce a hardened compact binary that can merge within reasonable timescales. We also stress that such a sharp decline is not dependent on the specific galactic prescriptions nor scatter in the adopted relations, but it is instead common to any approach including the production and possible merging of only isolated BH binaries. A viable solution is explored in the next Section.

\section{Discussion}\label{sec|discussion}

In this Section we discuss three interesting issues that could potentially affect our results: the impact of the adopted modeling prescriptions; the dynamical formation channel of BH binaries in dense environment such as star clusters; the implication of our work in the formation of light (super)massive BH seeds at high redshift.

\subsection{Impact of modeling prescriptions}\label{sec|discussion_modeling}

We warn the reader that the stellar BH relic mass function derived in the present work is somewhat dependent on the prescriptions entering the galactic and the stellar terms discussed in Sect. \ref{sec|galacticterm} and \ref{sec|stellarterm}. It is beyond the scope of this paper to address the issue in detail by exploring the variety of modeling recipes, numerical approaches, and associated parameter space present in the literature.  However, to provide the reader with a grasp of the related impact on our main results, we discuss the main source of uncertainties and show their impact on the local BH mass function.

As to the galactic term, the main source of uncertainty is constituted by the adopted main sequence and fundamental metallicity relationships. To estimate the related degree of uncertainty, we recomputed the galactic term by using in turn the main sequence relation by Boogaard et al. (2018) in place of our reference by Speagle et al. (2014), and the fundamental metallicity relation by Hunt et al. (2016) in place of our reference by Mannucci et al. (2011). The results are illustrated in Fig. \ref{fig|BHMF_galactic_comp}, and it is seen that the related changes on the stellar BH relic mass function are minor. 

As to the stellar term, the main source of uncertainty is constituted by the numerical treatment in the \texttt{SEVN} code of stellar winds, of pair-instability and pulsational pair-instability supernovae, of supernova explosions and natal kicks, of mass transfers and common-envelope phase in binary systems; all these physical processes can have profound impact on the formation and evolution of isolated or binary stellar BHs (see reviews by Postnov \& Yungelson 2006; Ivanova et al. 2013; Mapelli 2020 and references therein). To estimate the related degree of uncertainty, we recomputed the stellar term for binaries/failed binaries and the associated relic BH mass function by running a different binary stellar evolution code. We rely on \texttt{COSMIC} (Breivik et al. 2020) since it constitutes a state-of-the-art, community-developed software with extensive online documentation on the Github repository (see \url{https://cosmic-popsynth.github.io/}), that upgrades the classic and widespread code \texttt{BSE} (Hurley et al. 2002). As for the parameters regulating mass transfer, common envelope, tidal evolution, SN explosion and compact-object birth kicks, we keep the same choices adopted for \texttt{SEVN}, which are described in Spera et al. (2019). The results are presented in Fig. \ref{fig|BHMF_stellar_comp}. The top left and middle panels compare the (total) stellar term from the two codes. It is apparent that the main difference involves the distribution of remnants with mass $m_\bullet\gtrsim 50-60\, M_\odot$ that are associated to compact binaries and especially to failed BH binaries (i.e., binary stars have coalesced before evolving into two distinct binary BHs; see also Sect. \ref{sec|stellarterm}). Specifically, \texttt{COSMIC} tends to produce less failed BH binaries than \texttt{SEVN}; this is related to the treatment of the common envelope stage, which is indeed one of the most uncertain processes in the evolution of binary stars. The bottom panel illustrates the effects on the relic stellar BH mass function at $z\sim 0$. All in all, the differences are minor out to $m_\bullet \sim 60\, M_\odot$. At larger masses, the mass function built from the \texttt{COSMIC} stellar term decreases rapidly, to imply paucity of remnants with masses $m_\bullet\gtrsim 100\, M_\odot$; contrariwise, the mass function built from the \texttt{SEVN} stellar term is decreasing mildly and actually features a bump around $m_\bullet\sim 150\, M_\odot$, before a cutoff. Anyway, we stress that the major differences occur in a range where the mass function ${\rm d}N/{\rm d} m_\bullet$ is already decreasing much faster than $m_\bullet^{-1}$. 

Among the many parameters entering binary evolution codes, a major role is played by the common envelope  parameter $\alpha_{\rm CE}$, that quantifies the energy available to unbind the envelope (see Ivanova et al. 2013). As in Spera et al. (2019), we have set $\alpha_{\rm CE}=5$ throughout this paper and in the above comparison between \texttt{SEVN} and \texttt{COSMIC}. To have a grasp on the impact of such a choice, we have also run \texttt{COSMIC} with $\alpha_{\rm CE}=1$, and show in Fig. \ref{fig|BHMF_stellar_comp} the resulting stellar term (top right panel) and BH mass function at $z=0$ (bottom panel, dashed lines). The impact of $\alpha_{\rm CE}$ is rather limited, and mainly affects the high-mass tail of the distribution which is dominated by binary stellar evolution.

Another crucial parameter entering the stellar term is the binary mass fraction $f_{\star\star}$, for which we have assumed a constant value $0.5$. However, this is a rather uncertain quantity, that may well depend on the star mass $m_\star$, on the properties of the host galaxy and/or of the local environment, and even on redshift. In fact, observational constraints (e.g., Raghavan et al. 2010; Sana et al. 2012; Li et al. 2013; Sota et al. 2014; Dunstall et al. 2015; Luo et al. 2021) suggest values in the range $f_{\star\star}\approx 0.3-0.7$, with a possible increase toward more massive stars. To bracket the effects of different binary fractions on the stellar term and on the relic BH mass function, in Fig. \ref{fig|BHMF_fbin} we compare the results for our reference $f_{\star\star}=0.5$ to the extreme cases $f_{\star\star}=0$ and $f_{\star\star}=1$. Plainly adopting $f_{\star\star}=0$ (i.e., no stars born in binaries) cuts the tail of the stellar term and of the BH mass function for $m_\bullet\gtrsim 50-60\, M_\odot$, that are mostly produced by binary stellar evolution effects; meanwhile, for smaller masses the mass function is increased by the relatively more numerous single stars. Contrariwise, the other extreme case $f_{\star\star}=1$ (all stars born in binaries), tend to enhance the high-mass tail of the mass function and reduce somewhat the number density of BHs with masses $m_\bullet\lesssim 50-60\, M_\odot$. All in all, it is seen that the differences on the mass function are minor, and especially so for $f_{\star\star}\gtrsim 0.5$.

A final caveat about the IMF is in order. Throughout the paper we have self-consistently adopted as a reference the Kroupa (2001) IMF, mainly because it constitutes a standard both in the galaxy formation and the stellar evolution communities (together with the Chabrier 2003 IMF, which anyway would yield almost indistinguishable results) and because we had prompt availability of stellar evolution simulations based on that. Other classic choices like the Salpeter (1955), the Kennicutt (1983) and the Scalo (1986) IMF differ somewhat for the relative amount of star formation occurring below and above $1\, M_\odot$; such a difference is exacerbated in bottom-heavy IMFs like the one proposed to apply in massive ellipticals (e.g., van Dokkum \& Conroy 2010) or top-heavy IMFs like the one proposed to apply in the early Universe (e.g., Larson 1998) or in starburst galaxies (e.g., Lacey et al. 2010). It is important to stress that the IMF enters non-trivially both in the galactic and in the stellar term. In the galactic term, an IMF is needed to convert the observed galaxy luminosity functions into a statistics based on an intrinsic physical quantity such as the SFR or the stellar mass; moreover, the determination itself of SFR, stellar masses (hence of the main sequence of star-forming galaxies), star formation history and metallicity via broadband SED fitting rely on the assumption of a specific IMF.  Typically, at given SFR more top-heavy IMFs are proportionally richer in massive short-lived stars, and tend to yield a larger restframe UV luminosity/ionizing power, a faster chemical enrichment, and a smaller stellar mass locked in long-lived stars. In the stellar term, the IMF is plainly an essential ingredient in determining the relative proportion of compact remnants of different masses (cf. Eq. \ref{eq|singlestar}). More top-heavy IMFs tend to produce more massive stars per unit SFR, thus in principle more massive BH remnants (and a smaller relative amount of neutron stars), though this is not so straightforward due to mass loss and mass transfers during binary evolution. From the above considerations one could be led to speculate that precision determinations of the BH mass function in an extended mass range would constitute a probe for the IMF; however, this plan, at least presently, struggles against the many and large uncertainties associated with the treatment of binary evolution effects, and against the unexplored degeneracies with the bunch of poorly constrained parameters of galaxy and stellar evolution. 

\subsection{Dynamical channel}\label{sec|discussion_dynamics}

In Fig. \ref{fig|BHMF_primary} we highlighted a possible mismatch at $m_\bullet \gtrsim 40\, M_\odot$ between the primary mass distribution for merging BH binaries from isolated binary evolution and the estimates from gravitational wave observations. A viable solution could be that such large primary masses are produced in binary systems formed within the dense environment of young stellar clusters, open clusters, globular clusters, or nuclear star clusters (e.g., Di Carlo et al. 2019, 2020; Rodriguez et al. 2015, 2021; Antonini \& Rasio 2016; Kumamoto et al. 2019; Arca-Sedda et al. 2020; Banerjee 2021; Mapelli et al. 2021; Natarajan 2021). The central density of a star cluster can be so high that the orbits of binary stars are continuously perturbed by dynamical encounters with other members. Massive BHs $m_\bullet\gtrsim 40\, M_\odot$ in the pair instability mass gap can then be originated by hardening of BH binaries via dynamical exchanges in three-body encounters, and via the merging of massive progenitor stars; in addition, runaway collisions (i.e., a fast sequence of mergers; e.g., Portegies Zwart et al. 2004; Giersz et al. 2015; Mapelli 2016) in the densest cores of clusters with low metallicity can even produce intermediate mass BHs with $m_\bullet\gtrsim$ some $10^2\, M_\odot$.

To have a grasp on these effects, we proceed as follows. First, we construct the stellar term of Eq. (\ref{eq|stellarterm}) from the simulations by Di Carlo et al. (2020), that include dynamical effects in young star clusters. With respect to isolated conditions, an appreciable number of merging binaries with a primary mass $m_\bullet\gtrsim 40\, M_\odot$ is originated via dynamical exchanges, especially at low metallicities. More in detail, BHs with mass $m_\bullet\sim 40-65\, M_\odot$ can be formed (even in the field) from the evolution of single stars or stars in loose binaries that retain a fraction of their envelope up to their final collapse. BHs with mass $m_\bullet\gtrsim 65\, M_\odot$ can be originated via collisions of massive stars (eased in star clusters because dynamical encounters can induce fast merging before mass transfer episodes peel-off the primary star). If produced in the field, BHs from both these channels will remain isolated or locked in loose binaries unable to merge; contrariwise, in a star cluster they can acquire a new companion via dynamical exchanges and merge by dynamical hardening and gravitational wave emission. Note that, in contrast, repeated mergers of BHs are suppressed in young star clusters, because of their low escape velocity.

Second, we assume that a fraction $f_{\rm field}$ of the star formation occurs in the field and the complementary fraction $1-f_{\rm field}$ occurs in young star clusters (actually most of the stars are formed in young star clusters, but only a fraction of these may be subject to dynamical effects before exiting from the cluster or before the star cluster itself dissolves). Observations (see Goddard et al. 2010; Johnson et al. 2016; Chandar et al. 2017; Adamo et al. 2020) and cluster formation models (Kruijssen 2012; Pfeffer et al. 2018; Elmegreen 2018; El-Badry et al. 2019; Grudic et al. 2021) indicate that such a fraction is highly uncertain and possibly dependent on properties like the SFR spatial density and redshift; in this exploratory computation, we let the fraction $f_{\rm field}$ vary from $0.2$ to $1$ (which corresponds to isolated binaries only), and we split the galactic term of Eq. (\ref{eq|galacticterm}) accordingly. Finally, we combine the stellar and galactic term so derived to compute the merger rate of Eq. (\ref{eq|mergrate}) and the expected primary mass distribution. 

The outcome is illustrated in the top panel of Fig. \ref{fig|BHMF_starclusters} for different values of $f_{\rm field}$ in the range $0.2$ to $1$. As expected, increasing the fraction of SFR in star clusters (i.e., decreasing $f_{\rm field}$) produces a progressively more extended tail toward high primary masses, to the point that values $f_{\rm field}\lesssim 0.8$ actually can reconcile the theoretical prediction with the observational estimates. For completeness, in the bottom panel of Fig. \ref{fig|BHMF_starclusters} we show how the dynamical evolution channel affects the relic BH mass function at $z\sim 0$. The marked difference with respect to the model with only isolated binaries is the absence of the drop at around $m_\bullet\sim 60\, M_\odot$ and of the abrupt cutoff for $m_\bullet\sim 150\, M_\odot$. Instead, the mass function declines smoothly for $m_\bullet\gtrsim 60\, M_\odot$. The analytical fits in terms of Eq. (\ref{eq|fits}) for the field+cluster mass function with $f_{\rm field}=0.6$ at representative redshifts are reported in Table \ref{table|fits}.

Two caveats are in order here. First, in the simulations by Di Carlo et al. (2019, 2020) the treatment of stellar mergers is based on simplified assumptions: no mass-loss and chemical mixing during the merger; instantaneous recovery of hydrostatic equilibrium after the merger; rejuvenation of the merger product according to the simple Hurley et al. (2002) prescriptions. Plainly, these details of the process are quite uncertain, and dedicated hydrodynamical simulations are required to have a better understanding of the final outcome. Second, in estimating the impact of the dynamical formation channel we have considered only young star clusters for the sake of simplicity (and because the prompt availability of in-house dynamical simulations, that are extremely time-demanding to run from scratch). Globular clusters and nuclear star clusters could also be effective environments to build up massive binary BHs, since hierarchical mergers are more efficient in very rich and compact stellar systems (e.g., Miller \& Hamilton 2002; Antonini et al. 2019; Mapelli et al. 2021). Hence, including models for globular and nuclear star clusters could allow to reproduce the observed primary BH mass function with an even larger value of $f_{\rm field}$.

\subsection{BH seeds at high redshift}\label{sec|discussion_seeds}

We stress that the stellar BH mass function at high redshift $z\gtrsim 6$ derived in the present paper actually provides a light BH seed distribution in primordial galaxies, as originated by stellar and binary evolution processes. This is complementary to the seed distributions expected from other classic formation channels. The most relevant are basically three (see review by Volonteri 2010 and references therein): BHs formed by Pop III stars in metal-free environments like high-redshift minihalos at $z\gtrsim 20$ (e.g., Madau \& Rees 2001); runaway stellar collisions in metal-poor nuclear star clusters (e.g., Devecchi et al. 2012); direct collapse scenarios (e.g., Lodato \& Natarajan 2006). 

The light seed distributions from these models (extracted from Fig. 4 of Volonteri 2010) are compared in Fig. \ref{fig|BHMF_seeds} with the mass function at $z\sim 6-10$ from stellar BHs presented in this work. Given the appreciable contribution of the latter for masses $m_\bullet\lesssim 150\, M_\odot$ at redshift $z\lesssim 10$, it will be extremely relevant to include it in numerical simulations, semi-analytic and semi-empirical models of BH formation and evolution at high redshift.

\section{Summary}\label{sec|summary}

In this work we have provided an ab-initio computation of the relic stellar black hole (BH) mass function across cosmic times.  To this purpose, we have exploited the state-of-the-art stellar and binary evolutionary code \texttt{SEVN}, and have coupled its outputs with redshift-dependent galaxy statistics and empirical scaling relations involving galaxy metallicity, star-formation rate and stellar mass. Our main findings are summarized below.

\begin{figure}
\centering
\includegraphics[width=\textwidth]{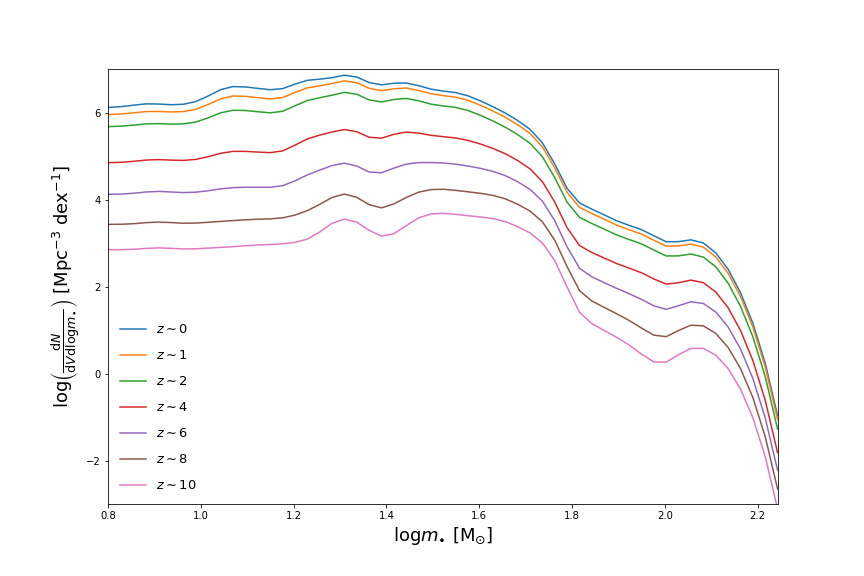}
\caption{The stellar BH relic mass function ${\rm d}N/{\rm d}V\,{\rm d}\log m_\bullet$ as a function of the BH mass $m_\bullet$ at different redshifts $z\sim 0$ (cyan), $z\sim 1$ (orange), $z\sim 2$ (green), $z\sim 4$ (red), $z\sim 6$ (violet), $z\sim 8$ (brown) and $z\sim 10$ (pink).}\label{fig|BHMF}
\end{figure}

\begin{figure}
\centering
\includegraphics[width=\textwidth]{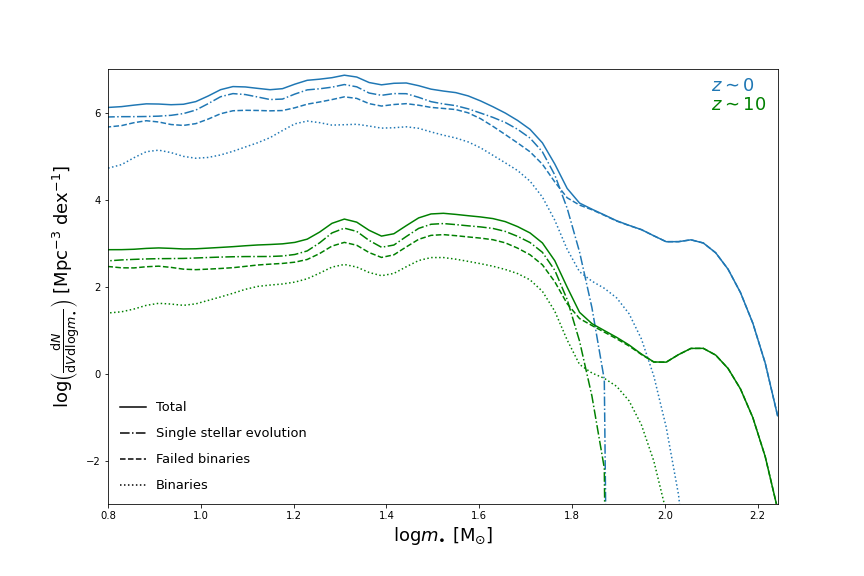}
\caption{The stellar BH relic mass function ${\rm d}N/{\rm d}V\,{\rm d}\log m_\bullet$ as a function of the BH mass $m_\bullet$ at redshift $z\sim 0$ (blue) and $z\sim 10$ (green), with highlighted the contribution to the total (solid lines) by single BHs formed from isolated stars (dot-dashed lines), failed BH binaries (dashed lines), and BH binaries (dotted lines).}\label{fig|BHMF_split}
\end{figure}

\begin{figure}
\centering
\includegraphics[width=\textwidth]{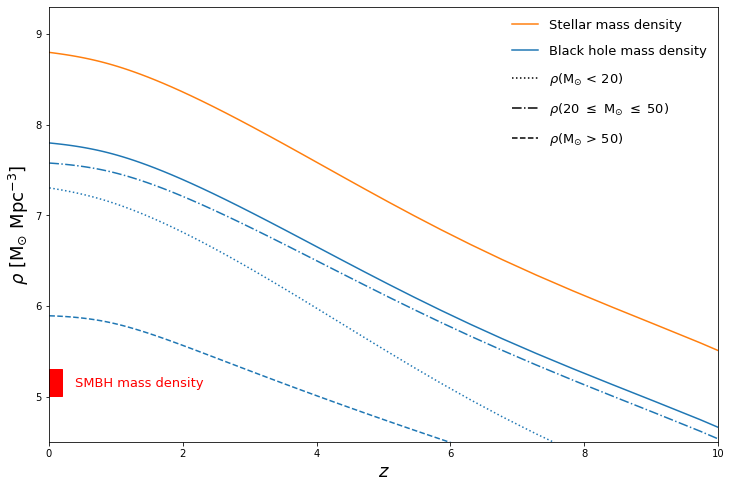}
\caption{The relic density $\rho_\bullet$ in stellar mass BHs as a function of redshift $z$ (solid blue line); contribution from different BH mass ranges are also displayed: $m_\bullet/M_\odot\lesssim 20$ (dotted), $20\lesssim m_\bullet/M_\odot\lesssim 50$ (dashed) and $m_\bullet/M_\odot\gtrsim 50$ (dot-dashed). For reference, the stellar mass density in galaxies is also illustrated (orange line).
Moreover, the observational estimate of the mass density in supermassive BHs with $M_\bullet\sim 10^6-10^{10}\, M_\odot$ at $z\sim 0$ is shown (Shankar et al. 2020; red shaded area in the bottom left corner).}\label{fig|BHMF_density}
\end{figure}

\begin{figure}
\centering
\includegraphics[width=\textwidth]{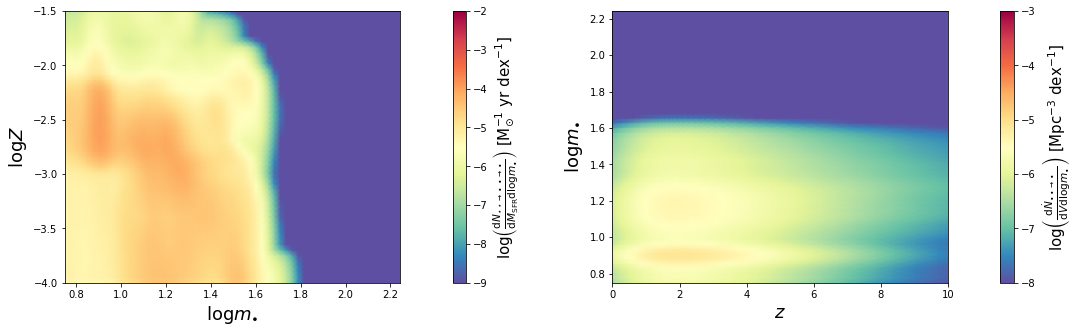}\\\includegraphics[width=\textwidth]{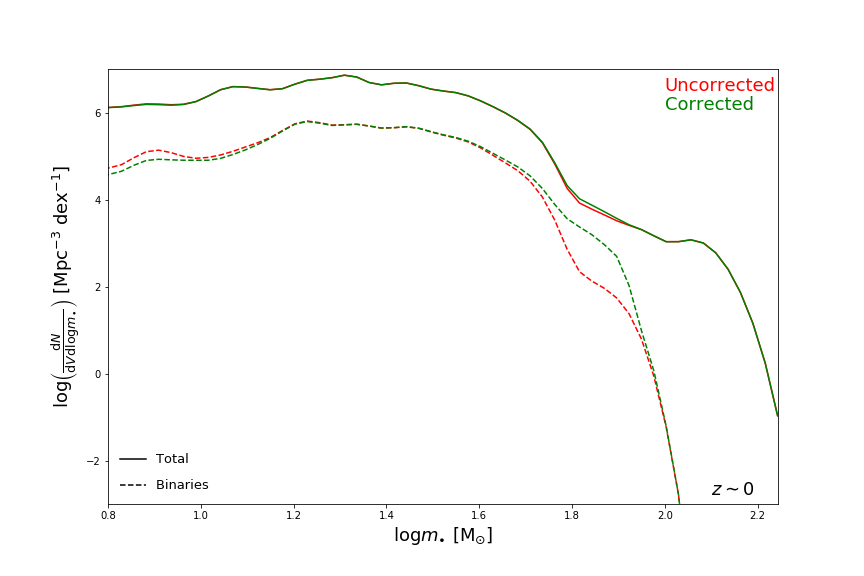}\caption{Impact of binary BH mergers on the stellar BH relic mass function. Top left panel: stellar term $\sum_{i=1,2}\,{\rm d}N_{\star\star\rightarrow\bullet\bullet\rightarrow \bullet}/{\rm d}M_{\rm SFR}\, {\rm d}\log m_{\bullet,i}$ representing the number of binary stellar systems that first evolve in a BH binary and then are able to merge within the Hubble time (color-coded), as function of metallicity $Z$ and BH mass $m_\bullet$. Top right panel: cosmic merging rate $\sum_{i=1,2}\,{\rm d}\dot N_{\bullet\bullet\rightarrow \bullet}/{\rm d}V\, {\rm d}\log m_{\bullet,i}$ (color-coded; see Eq. \ref{eq|mergrate}) as a function of BH mass $m_\bullet$ and redshift $z$. Bottom panel: stellar BH relic mass function ${\rm d} N_{\bullet}/{\rm d}V\, {\rm d}\log m_{\bullet}$ as a function of BH mass $m_\bullet$ at redshift $z\sim 0$, without (red lines) and with (green line) correction for binary BH merging effects; solid lines refer to the total mass function and dashed line to the contribution from binary BHs.}\label{fig|BHMF_merging}
\end{figure}

\begin{figure}
\centering
\includegraphics[width=\textwidth]{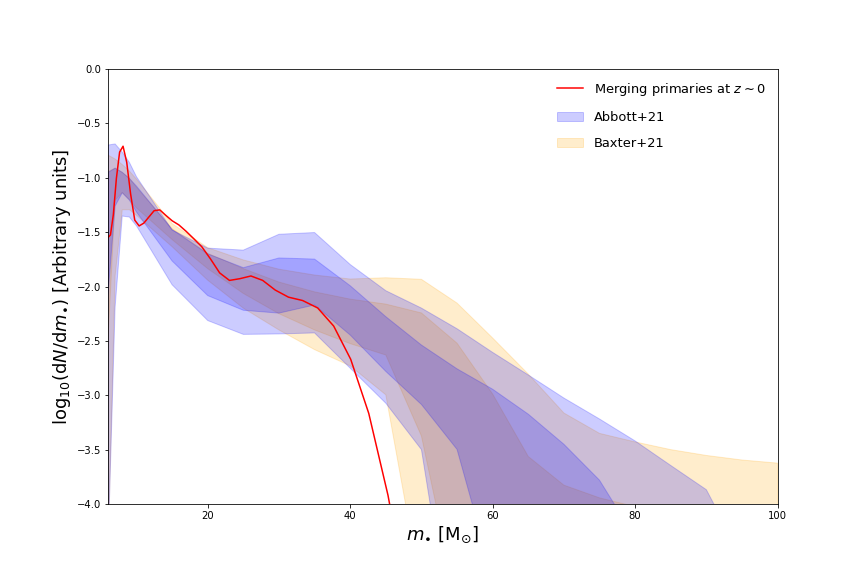}\caption{The BH mass function of merging BH binaries as a function of primary BH mass at $z\approx 0$. The outcome of this work is illustrated by the red solid line. Estimates from the analysis of gravitational wave observations by Abbott et al. (2021a) and Baxter et al. (2021) are reported as blue and orange shaded areas, respectively (for both determinations the $68\%$ and $95\%$ credible intervals are represented with dark and light shades).}\label{fig|BHMF_primary}
\end{figure}

\begin{figure}
\centering
\includegraphics[width=\textwidth]{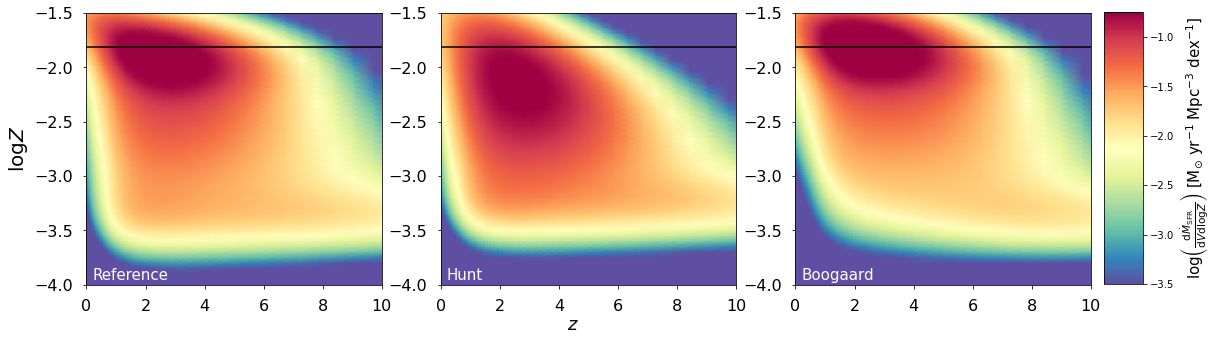}\\\includegraphics[width=\textwidth]{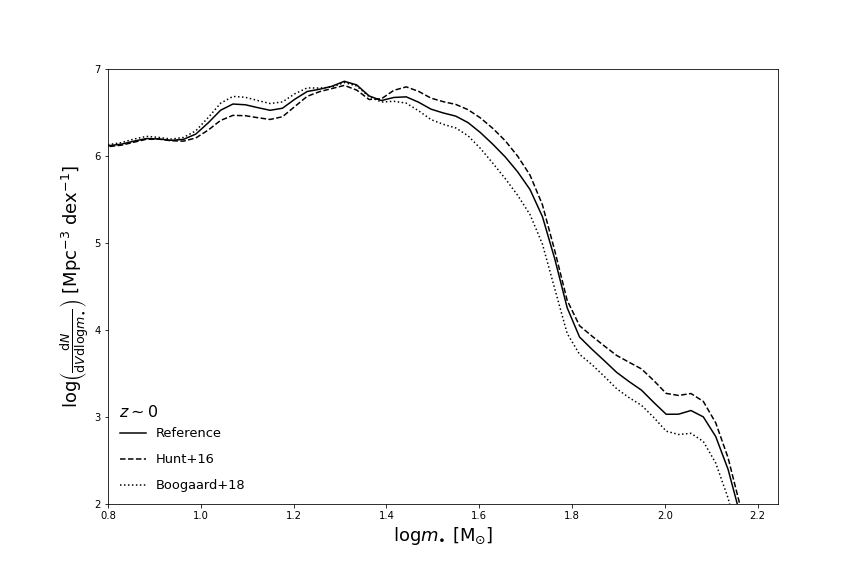}\caption{Impact on the BH mass function of different prescriptions for the galactic term (see Sect. \ref{sec|discussion_modeling}). Top panels: the galactic term of Eq. (\ref{eq|galacticterm}) when assuming
our reference Speagle (2014) main-sequence and Mannucci et al. (2011) fundamental metallicity relation (left), when changing the fundamental metallicity relation to that by Hunt et al. (2016; middle), and when changing the main sequence relation to that by Boogaard et al. (2018; right). Bottom panel: the relic stellar BH mass function at redshift $z\sim 0$ for the three cases just described above.}\label{fig|BHMF_galactic_comp}
\end{figure}

\begin{figure}
\centering
\includegraphics[width=\textwidth]{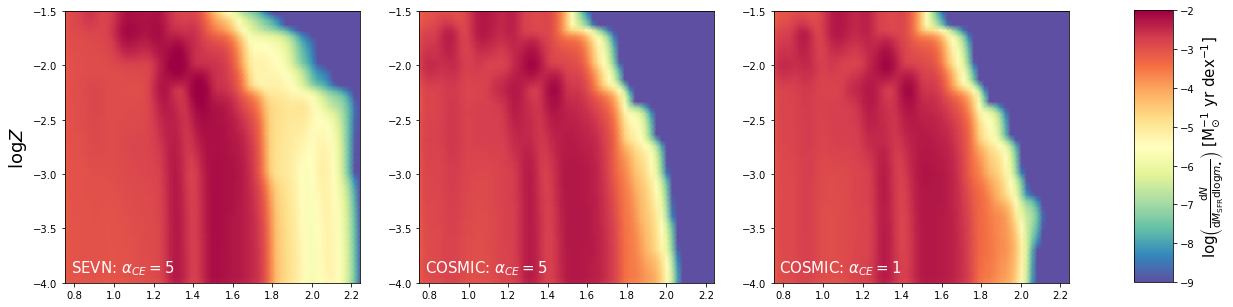}\\\includegraphics[width=\textwidth]{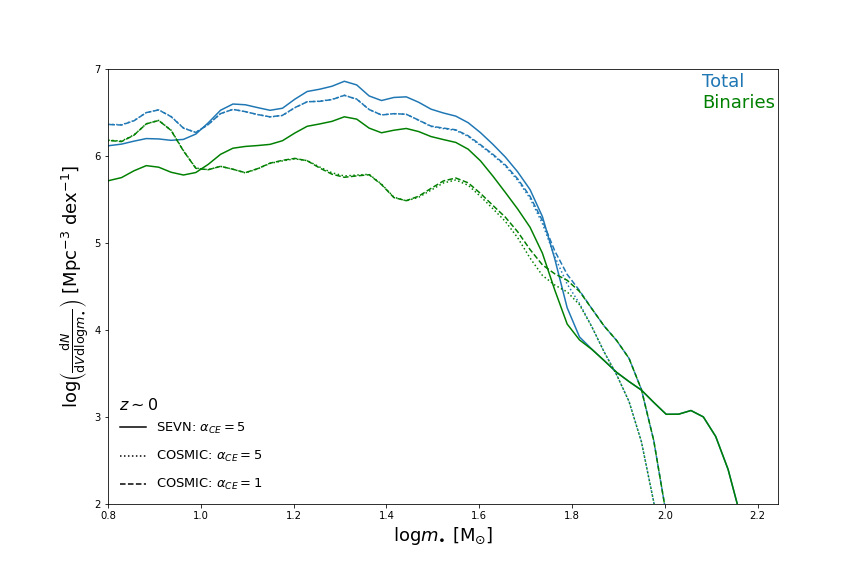}\caption{Impact on the BH mass function of different codes and prescriptions for the computation  of the stellar term (see Sect. \ref{sec|discussion_modeling}). Top panels: the (total) stellar term of Eq. (\ref{eq|stellarterm}) from the \texttt{SEVN} (left) and the \texttt{COSMIC} (middle) binary stellar evolution codes assuming a common envelope parameter $\alpha_{\rm CE}=5$ (see Sect. \ref{sec|discussion_modeling}); results with \texttt{COSMIC} (right) for $\alpha_{\rm CE}=1$ are also shown. Bottom panel: the relic stellar BH mass function at redshift $z\sim 0$ from the \texttt{SEVN} (solid blue line) and the \texttt{COSMIC} (dotted blue line) with $\alpha_{\rm CE}=5$, and from \texttt{COSMIC} (dashed blue line) with $\alpha_{\rm CE}=1$; the contribution from binaries is also highlighted (green lines).}\label{fig|BHMF_stellar_comp}
\end{figure}

\begin{figure}
\centering
\includegraphics[width=\textwidth]{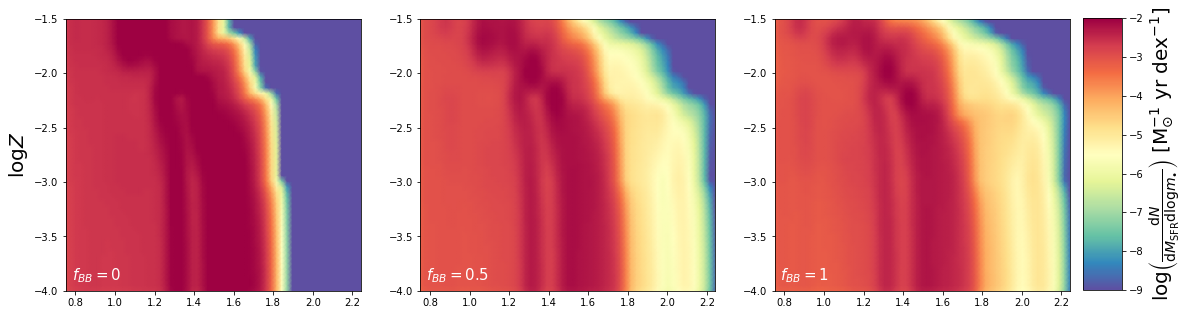}\\\includegraphics[width=\textwidth]{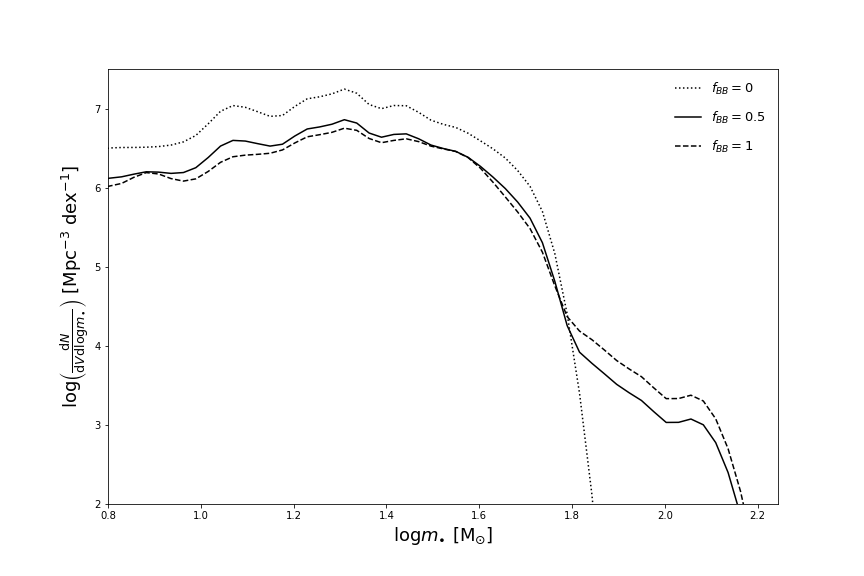}\caption{Impact of the binary fraction $f_{\star\star}$ on our results (see Sect. \ref{sec|discussion_modeling}). Top panels: the (total) stellar term of Eq. (\ref{eq|stellarterm}) for $f_{\star\star}=0$ (left; no stars born in binaries), $0.5$ (middle; our fiducial case) and $1$ (right; all stars born in binaries). Bottom panel: the relic stellar BH mass function at redshift $z\sim 0$ for $f_{\star\star}=0$ (dotted line), $0.5$ (solid) and $1$ (dashed).}\label{fig|BHMF_fbin}
\end{figure}

\begin{figure}
\centering
\includegraphics[width=0.8\textwidth]{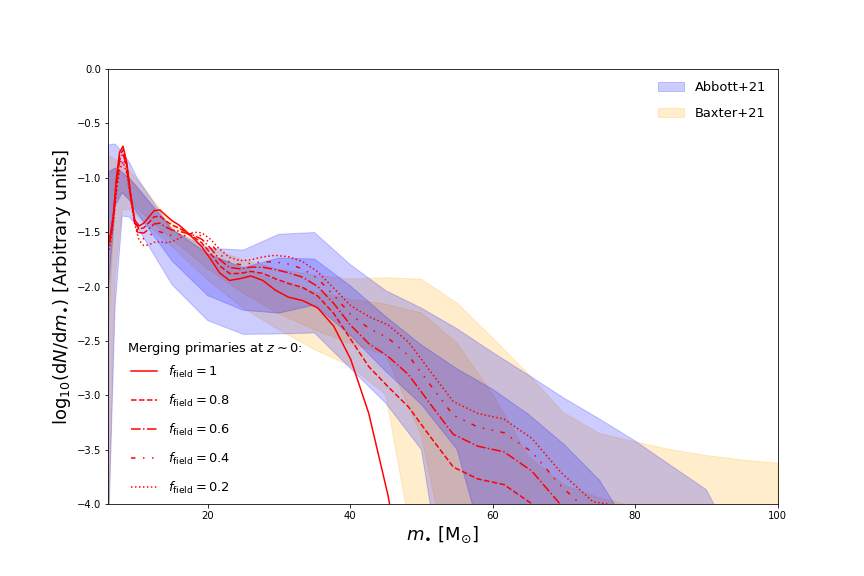}\\\includegraphics[width=0.8\textwidth]{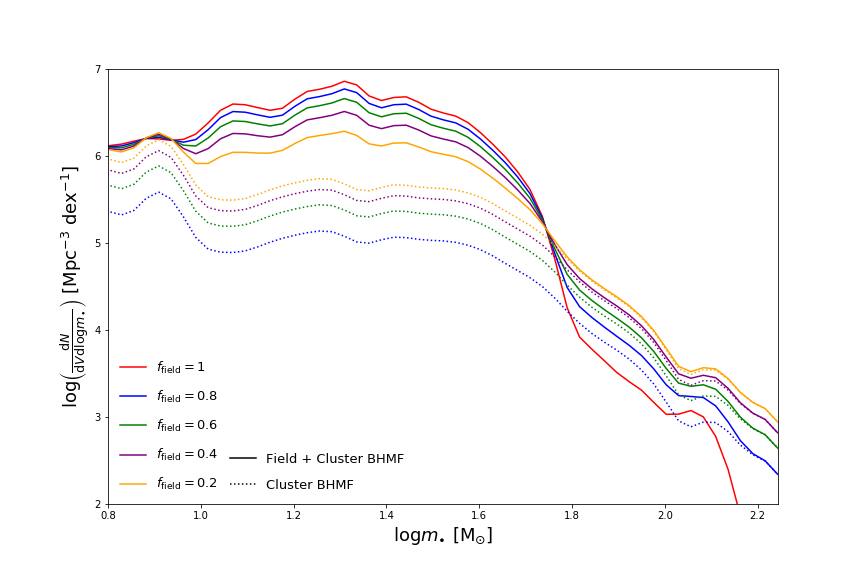}\caption{Top panel: impact of the dynamical formation channel on the BH primary mass distribution of merging BH binaries $z\approx 0$, computed from the simulations for young open star clusters by Di Carlo et al. (2020, see Section \ref{sec|discussion_dynamics} for details). Lines refers to different fraction of star formation occurring in the field $f_{\rm field}\approx 1$ (solid; only isolated binaries), $0.8$ (dashed), $0.6$ (dot-dashed), $0.4$ (dot-dot-dashed) and $0.2$ (dotted). Estimates from the analysis of gravitational wave observations by Abbott et al. (2021a) and Baxter et al. (2021) are reported as blue and orange shaded areas, respectively (for both determination the $68\%$ and $95\%$ credible intervals are represented with dark and light shades). Bottom panel: impact of the dynamical formation channel on the relic BH mass function at $z\sim 0$. Dotted lines refer to the contribution from star clusters and solid lines to the total (field+star clusters) BH mass function. Color-code refers to different fraction of star formation occurring in the field $f_{\rm field}\approx 1$ (red), $0.8$ (blue), $0.6$ (green), $0.4$ (purple) and $0.2$ (yellow).}\label{fig|BHMF_starclusters}
\end{figure}

\begin{figure}
\centering
\includegraphics[width=\textwidth]{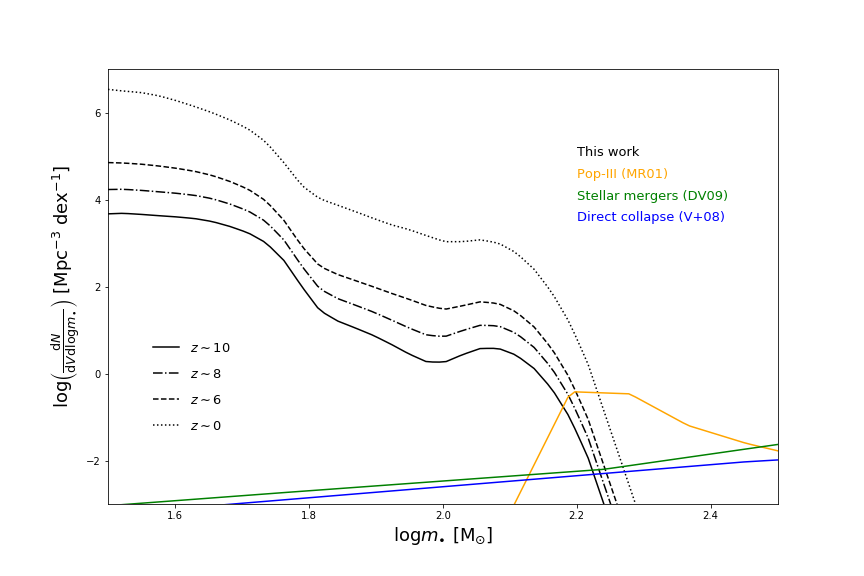}\caption{The light seed distribution at redshift $z\sim 10$ (solid), $8$ (dot-dashed), $6$ (dashed) and $0$ (dotted) from the stellar BH mass function of this work is compared with that expected from pop-III remnants (orange line), runaway stellar collisions in nuclear star clusters (green line), and direct collapse scenarios (blue line); see discussion in Sect. \ref{sec|discussion_seeds}. }\label{fig|BHMF_seeds}
\end{figure}

\begin{itemize}
    \item The relic mass function ${\rm d}N/{\rm d}V{\rm d}\log m_\bullet$ as a function of the BH mass $m_\bullet$ features a rather flat shape up to $m_\bullet\approx 50\, M_\odot$ and then a log-normal decline for larger masses, while its normalization increases with decreasing redshift, quite rapidly down to $z\sim 2-3$ and then more mildly toward $z\sim 0$, see Fig. \ref{fig|BHMF} and Table \ref{table|fits}. The local stellar BH mass function could be eventually probed via 
    microlensing observations (see Paczynski 1986; for a review, Mao 2012).
    
    \item The local relic BH mass function for $m_\bullet\lesssim 50\, M_\odot$ is comparably contributed from isolated stars evolving into BH and from binary stellar systems ending up in single BH (failed BH binaries), while binary BHs are subdominant; for higher masses $m_\bullet\gtrsim 50\, M_\odot$ the single stellar evolution and binary BH channels abruptly decrease, while the failed BH binaries dominates largely. See Fig. \ref{fig|BHMF_split}. 
      
    \item The local stellar BH relic mass density amounts to $\rho_\bullet\approx 5\times 10^7\, M_\odot$ Mpc$^{-3}$, exceeding by more than two orders of magnitude that in supermassive BHs; this translates into an energy density parameter $\Omega_\bullet\approx 5\times 10^{-4}$, implying that the total mass in stellar BHs amounts to $\lesssim 1\%$ of the local baryonic matter. See Fig. \ref{fig|BHMF_density}.
    
    \item The stellar BH relic mass function can be distorted by binary  BH mergers, that can redistribute remnants from the low to the high mass range. However, such a reshaping is found to have a minor effect on the mass function at the high-mass end. See Fig. \ref{fig|BHMF_merging}. 
 
    \item{}The distribution of primary masses for merging BH binaries is found to be in remarkable agreement with the  recent estimates from gravitational wave observations by LIGO/Virgo out to $m_\bullet\lesssim 40\, M_\odot$. For larger masses, the observed distribution declines gently while the theoretical one dies off due to a twofold reason: the mass gap from pair-instability and pulsational pair-instability supernovae; and the substantial mass loss during the common envelope phase needed to produce a tight BH binary that can merge within the Hubble time. We have proposed as a viable solution to consider the dynamical formation of merging BH binaries in dense environment like (young) star clusters. See Figs. \ref{fig|BHMF_primary}-\ref{fig|BHMF_starclusters} and Table \ref{table|fits}. 
    
    \item{}We have discussed the impact on the mass function of adopting different physical prescriptions entering the galactic and the stellar term. As for the galactic term, minor differences arise when changing the adopted main sequence of star-forming galaxies and the fundamental metallicity relationship. As for the stellar term, the main differences concern the high mass end of the mass function, and are due to the numerical treatment of binary stellar evolution effects, which are still considerably uncertain even from a theoretical point of view. See Figs. \ref{fig|BHMF_stellar_comp} and \ref{fig|BHMF_galactic_comp}.
    
    \item The BH mass function derived here can provide a firm theoretical basis for a physically-motivated light seed distribution at high redshift; as expected, for masses $m_\bullet\lesssim 150\, M_\odot$ and redshifts $z\lesssim 10$ it overcomes complementary light seed formation channels, like Pop-III stars, stellar mergers in nuclear star clusters and direct collapse scenarios that are aimed to produce larger mass seeds. It will be worth implementing the light seed distribution from the present work in semi-analytic and numerical models of (super)massive BH formation and evolution, as it can be quite relevant for predictions concerning future gravitational wave observations via Einstein Telescope and LISA. See Fig. \ref{fig|BHMF_seeds}.
    
\end{itemize}

In a future perspective, it would be interesting to exploit the galactic and stellar evolution prescriptions adopted here to populate a $N-$body simulation via subhalo clustering abundance matching technique (e.g., Ronconi et al. 2020) in order to derive a mock catalog encapsulating the three-dimensional spatial distribution of stellar-mass BHs and of their galactic hosts. Finally, we stress that our work can constitute a starting point to investigate the origin of heavy seeds and the growth of (super)massive BHs in high-redshift star-forming galaxies, that we will pursue in forthcoming papers.

\acknowledgements

We thank the referee for a competent and constructive report. We acknowledge C. Baccigalupi and M. Massardi for interesting discussions and critical reading. This work has been supported by the EU H2020-MSCA-ITN-2019 Project 860744 “BiD4BESt: Big Data applications for black hole Evolution STudies.” AL acknowledges funding from the PRIN MIUR 2017 prot. 20173ML3WW, “Opening the ALMA window on the cosmic evolution of gas, stars and supermassive black holes”. UNDC and MM acknowledge financial support from the European Research Council for the ERC Consolidator grant DEMOBLACK, under contract no. 770017.

\newpage

\newpage

\newpage

\newpage

\newpage

\newpage

\newpage

\clearpage

\clearpage

\clearpage

\clearpage

\clearpage

\clearpage

\clearpage

\clearpage
\begin{deluxetable}{lccccccccccccccccccccc}
\tablewidth{0pt}\tablecaption{Fits to stellar BH relic mass function via Eq. (\ref{eq|fits}).}\tablehead{& &\multicolumn{6}{c}{Field, $f_{\rm field}=1$} & & \multicolumn{6}{c}{Field+Cluster, $f_{\rm field}=0.6$}\\ \cline{3-8} \cline{10-15}\\
 \colhead{$z$} & & \colhead{$\log\mathcal{N}$} &  \colhead{$\log\mathcal{M}_\bullet$} &  \colhead{$\alpha$}  & 
 \colhead{$\log \mathcal{N}_G$} & $ \log \mathcal{M}_{\bullet, G}$ &
 \colhead{$\sigma_G$}  & & \colhead{$\log\mathcal{N}$} &  \colhead{$\log\mathcal{M}_\bullet$} &  \colhead{$\alpha$}  & 
 \colhead{$\log \mathcal{N}_G$} & $\log \mathcal{M}_{\bullet, G}$ &
 \colhead{$\sigma_G$}\\ & &  [Mpc$^{-3}$] &  $[M_\odot]$ & &  [Mpc$^{-3}$] & $[M_\odot]$ & & & [Mpc$^{-3}$] &  $[M_\odot]$ & &  [Mpc$^{-3}$] & $[M_\odot]$ & &}
 \startdata
        0 & & 5.623  & 0.607  & -3.781  & 2.413  & 2.021 & 0.052 & & 6.078 & 0.704 & -2.717 & 3.496 & 1.808 & 0.1846\\
        1 & & 5.429 & 0.609 & -3.859 & 2.309 & 2.023 & 0.051 & 
        & 5.887 & 0.709 & -2.785 & 3.304 & 1.843 & 0.173\\
        2 & & 5.107 & 0.612 & -3.914 & 2.064 & 2.024 & 0.051 & 
        & 5.592 & 0.713 & -2.823 & 3.008 & 1.866 & 0.165\\
        4 & & 4.344 & 0.634 & -3.902 & 1.419 & 2.037 & 0.049 & 
        & 4.796 & 0.747 & -2.782 & 2.101 & 1.952 & 0.132\\
        6 & & 3.614 & 0.659 & -3.866 & 0.806 & 2.054 & 0.045 &
        & 4.112 & 0.785 & -2.718 & 1.359 & 2.012 & 0.107\\
        8 & & 2.894 & 0.676 & -3.868 & 0.197 & 2.066 & 0.043 &
        & 3.457 & 0.816 & -2.660 & 0.685 & 2.046 & 0.091\\
        10 & &2.305 & 0.680 & -3.884 & -0.344& 2.072 & 0.042 & 
        & 2.897 & 0.831 & -2.623 & 0.113 & 2.059 & 0.0841\\
\enddata
\tablecomments{Fits valid in the range $m_\bullet\sim 5-160\, M_\odot$. Typical relative uncertainties on the parameters are $\lesssim 10\%$, and typical values of the reduced $\chi^2_r\lesssim 1$ are found. The results for $f_{\rm field}=0.6$ are based on the simulations by Di Carlo et al. (2019, 2020) for young open star clusters.}\label{table|fits}
\end{deluxetable}

\end{document}